\documentclass[aps,prd,reprint,superscriptaddress,nofootinbib, amsfonts]{revtex4-2}
\usepackage[all]{xy}
\usepackage{amsmath,amsthm,amssymb}
\usepackage[dvips]{graphicx}
\usepackage{comment}
\usepackage{array}
\usepackage{bm}
\allowdisplaybreaks[1]

\usepackage{xcolor}
\usepackage{hyperref}
\usepackage{tcolorbox}
\usepackage{mathrsfs}
\usepackage{float}
\usepackage[caption=false]{subfig}
\usepackage{orcidlink}

\newcommand{\rmin}{r_{\rm min}}

\newcommand{\rmax}{r_{\rm max}}

\newcommand{\vf}{\varphi}

\newcommand{\wmax}{\omega_{\rm max}}

\newcommand{\lm}{\ell m}
\newcommand{\lw}{\ell\omega}
\newcommand{\lmw}{\ell m \omega}

\newcommand{\SP}[1]{\left(#1\right)}

\newcommand{\SB}[1]{\left[#1\right]}

\begin{document}

%\title{Frequency-domain calculation of the scalar-field self-force in hyperbolic scattering}
\title{Frequency-domain self-force calculations using Gegenbauer reconstruction}

\def\Birmingham{Institute for Gravitational Wave Astronomy and School of Physics and Astronomy, University of Birmingham, Edgbaston, Birmingham, B15 2TT, United Kingdom}
\def\Soton{Mathematical Sciences, University of Southampton, 
Southampton SO17 1BJ, United Kingdom}
\def\AEI{Max Planck Institute for Gravitational Physics (Albert Einstein Institute), D-14476 Potsdam, Germany}

\author{Christopher Whittall \orcidlink{0000-0003-2152-6004}}
\email{c.l.whittall@bham.ac.uk}
\affiliation{\Birmingham}
\affiliation{\Soton}

\author{Leor Barack \orcidlink{0000-0003-4742-9413}}
\affiliation{\Soton}

\author{Oliver Long \orcidlink{0000-0002-3897-9272}}
\affiliation{\AEI}

\date{\today}
\begin{abstract} 
We investigate the use of the Gegenbauer procedure for time-domain reconstruction in frequency-domain calculations of the self-force on a particle orbiting a black hole. The conventional technique relies on the so-called method of extended homogeneous solutions (EHSs), which circumvents the Gibbs phenomenon that would otherwise hamper the reconstruction of the particle's field from frequency modes. Unfortunately, the effectiveness of EHS reconstruction deteriorates rapidly with increasing orbital eccentricity, due to large numerical cancellations between frequency-mode contributions. Furthermore, the method is only suitable for bound orbits and cannot be fully applied in scattering scenarios. The Gegenbauer reconstruction procedure involves a reprojection of the partial Fourier representation onto a complementary basis of polynomials. The resulting series converges exponentially while averting the cancellation problem. We demonstrate the procedure with numerical results for the spherical harmonic modes of the scalar field sourced by a scalar charge in orbit around a Schwarzschild black hole. We illustrate the merits of this approach, and discuss the challenges that remain for a full Gegenbauer-based self-force calculation.

\end{abstract}

\maketitle

%\tableofcontents

\section{Introduction}
Gravitational self-force is a perturbative approach to modeling the gravitational-wave-driven dynamics of compact binary systems in general relativity, applicable to systems where one object is significantly more massive than the other but without any restriction on the velocities or separation of the binary \cite{Barack:2018yvs, Pound:2021qin}. Self-force methods underpin the program to model extreme-mass-ratio inspirals as potential sources for gravitational-wave astronomy \cite{LISA_waveform_wp}, and they also play an important role in studies of the gravitational interaction in two-body scattering \cite{Damour2020,Gralla:2021qaf,BarackLong2022,Barack:2023oqp,Long:2024ltn,Cheung:2024byb}. In the self-force approach, the equations governing the orbital dynamics and emitted gravitational waves are systematically solved order by order in the small mass ratio $\mu/M$. In the limit $\mu/M\to 0$, the system is described by a test particle moving along a geodesic orbit in the geometry of the larger body, usually assumed to be a (Kerr) black hole.  At $O(\mu/M)$ the small body's gravitational field is treated as a linear perturbation of the background black hole geometry. The point-particle description still holds, but now backreaction from the perturbation produces an effective self-force that slowly accelerates the particle away from geodesic motion in the background geometry. 

Calculations of the self-force-driven binary evolution need as input (at least) the linear metric perturbation sourced by the particle. Unless further approximation assumptions are made, in general this perturbation field can only be obtained numerically, by solving the linearized field equations with a suitable delta-function source and suitable boundary conditions. This is usually done mode by mode in a multipole decomposition using a suitable basis of harmonics defined on spheres around the large black hole. In appropriate gauges, the modal field contributions are bounded even at the location of the particle, though they have a finite differentiability there, a legacy of the Coulomb-like singularity in the full field. 
The self-force can be computed directly from these modal contributions, and their derivatives at the particle's location, using the standard procedure of mode-sum regularization \cite{Barack:1999wf}. 

In a frequency-domain self-force calculation, the multipole modes of the perturbation field are further decomposed into Fourier modes, with the obvious advantage of reducing the numerical task to the solution of a set of ordinary differential equations. In bound-orbit problems the field exhibits a discrete Fourier spectrum, inherited from the periodic nature of the background Kerr geodesics, whereas in scattering calculations the spectrum is continuous and a Fourier integral is used instead. In either case, a complication arises when one attempts to reconstruct the time-domain field and its derivatives at the particle's location, due to the finite differentiability of the field there. A naive, direct reconstruction is rendered impractical by the Gibbs phenomenon, which disrupts the convergence of the frequency mode sum or integral (an illustration is provided in Fig.~\ref{fig:direct_vs_ehs_vs_td} below).

The standard workaround employs what is known as the method of {\it extended homogeneous solutions} (EHSs) \cite{Barack:2008ms}. The procedure is applicable in situations where the sourcing particle is confined to a compact range of radii outside the black hole, $r_{\rm min}\leq r_p(t)\leq r_{\rm max}$, as in the case of bound geodesic orbits. Here $r,t$ are Boyer-Lindquist coordinates on the background Kerr geometry, and $r=r_p(t)$ describes the particle's orbit.  In the EHS method, the time-domain field at the particle is reconstructed not from the true Fourier modes of the particle's field (i.e., solutions to the inhomogeneous, sourced frequency-domain field equations), but from certain homogeneous solutions, obtained by analytically extending the frequency modes from the vacuum regions $r<r_{\rm min}$ and $r>r_{\rm max}$ into the libration domain. The Fourier sum of the extended EHS modes recovers the true time-domain solution on $r\leq r_p(t)$ for the outwardly extended modes and on $r\geq r_p(t)$ for the inwardly extended ones, and both sums maintain exponential convergence. In the scattering case, where the radial extent of the source is only half compact,  $r_p(t)>r_{\rm min}$, the method can only be used to reconstruct the field on $r\leq r_p(t)$. The EHS method will be reviewed in more detail in Sec.\ \ref{sec:EHS_reconstruction}. It has been used extensively in calculations for bound orbits, and recently also for scattering \cite{Whittall:2023xjp}.   

However, the EHS method has a notable weakness. The problem was first identified in Ref.\ \cite{vandeMeent2016}, and further diagnosed and analyzed in \cite{Whittall:2023xjp}. The symptom is a rapidly increasing loss of numerical accuracy in the reconstructed time-domain field when extending far into the libration region. The root cause is that the EHS modes grow polynomially in $r$---and exponentially in $\ell$---as they are extended into the source region, a behavior that is not shared by the true solution. Hence, the recovery of the true solution in the source region involves a delicate cancellation between EHS frequency mode contributions. The maximum degree of cancellation encountered grows with eccentricity for a bound-orbit source, and the cancellation can become arbitrarily severe as $r\rightarrow \infty$ for a scattering source. In both cases, the cancellation grows exponentially with $\ell$. This fundamentally restricts the ability of finite-precision floating point numerical calculations based on EHSs to tackle large-eccentricity orbits or the large-$r$ portion of scattering orbits. In past implementations \cite{vandeMeent2016, vandeMeent2018} the problem  was somewhat mitigated via the use of arbitrary precision arithmetic.  But it remains beyond the reach of EHS-based calculations to tackle bound orbits with eccentricities much above $\sim 0.7$, or the portion of any scattering orbit at radii greater than a few dozen Schwarzschild radii above $r_{\rm min}$.

A further weakness of EHSs manifests itself in the scattering case. Since hyperbolic scattering orbits are not bounded from above by a vacuum region, EHSs cannot {\it a priori} be used for constructing the field in the region $r>r_p(t)$. One is restricted to obtaining the field and its one-sided derivatives at $r\to r_p^-$ only. A self-force calculation is still possible, but it is much more involved than when two-sided information is available \cite{Pound:2013faa}.

These inherent problems with EHSs motivate the question of whether an alternative method might be devised for dealing with the Gibbs phenomenon, one which completely does away with EHSs.

In this paper, we explore one such potential alternative. The proposed method circumvents the Gibbs phenomenon while avoiding the use of the EHS and its attendant cancellation problem. When applied to scattering orbits, it will allow us to reconstruct the field on both $r<r_p(t)$ and $r>r_p(t)$. The new approach is inspired by a familiar technique in applied mathematics, known as {\em Gegenbauer reconstruction}. The method relies on a reprojection of the partial Fourier representation onto a complementary basis of Gegenbauer polynomials (a generalization of both Legendre and Chebyshev polynomials), whose convergence properties have been demonstrated to reduce Gibbs artifacts \cite{gottlieb1992gibbs, gottlieb1994resolution, gottlieb1995gibbs, gottlieb1997gibbs}. This will allow us to accurately reconstruct the perturbation field from the frequency modes of the physical inhomogeneous field, with exponential convergence of the mode sum even near and at the particle. 

We demonstrate this approach with a numerical calculation in a scalar-field model for a scalar charge on a fixed scattering geodesic in the Schwarzschild spacetime. Specifically, we reconstruct individual spherical-harmonic modes of the scalar field, and their derivatives, on the worldline of the scalar charge. The scalar-field model we work with is introduced in Sec.~\ref{sec:preliminaries}. In Sec.~\ref{sec:FDmethods} we introduce the frequency-domain decomposition of the inhomogeneous scalar-field equation and its solution, before describing the method of the EHS and its cancellation problem. Section~\ref{sec:Geg_intro} then presents the Gegenbauer reconstruction procedure and its application in a toy analytical example. The procedure is applied in Sec.~\ref{sec:scalar_field} to the problem of a scalar field with a scattering geodesic source on a Schwarzschild background. For benchmarking we use numerical results from EHS reconstruction where possible [i.e., on $r\leq r_p(t)$ at small separations], or otherwise results from time-domain calculations \cite{BarackLong2022}. We particularly emphasize the ability of the Gegenbauer procedure to reconstruct the field on $r>r_p(t)$ and at large orbital separations, where the EHS approach fails. We conclude in Sec.~\ref{sec:discussion} with an analysis of the merits and limitations of the Gegenbauer procedure, and discuss remaining work necessary to enable full self-force calculations using this approach.

Throughout this work we use natural units with $G = 1 = c$. 

\section{Preliminaries: scalar-field self-force and scattering} \label{sec:preliminaries}

Our specific motivation for this study comes from self-force calculations for scattering orbits, and we will therefore develop and illustrate our method in this context. We will start, in this section and the next, by reviewing the frequency-domain approach to self-force calculations for a scalar charge on a scattering geodesic in Schwarzschild spacetime. We do this to set up the physical model, introduce notation, and highlight the necessary numerical input for such calculations.   

\subsection{Scattering geodesics in Schwarzschild spacetime}

We assume the primary object is a Schwarzschild black hole of mass $M$. Its geometry is described by the line element
\begin{align}
    ds^2 = -f(r)dt^2 + f(r)^{-1}dr^2 + r^2d(d\theta^2 + \sin^2\theta\, d\vf^2),
\end{align}
where $f(r) := 1-2M/r$. The smaller object is described by a pointlike particle carrying mass $\mu \ll M$ and scalar charge $q \ll \sqrt{\mu M}$. Its trajectory is described by a timelike worldline $x_p^{\alpha}(\tau)$ in the background spacetime, parametrized by proper time $\tau$, with 4-velocity $u^{\alpha}(\tau) := dx_p^{\alpha}/d\tau$. 

In the test-particle limit, $\mu/M \rightarrow 0$ with $q^2/(\mu M) \rightarrow 0$, the particle moves along a geodesic in the background Schwarzschild spacetime. This geodesic is taken to lie in the equatorial plane, $\theta = \pi/2$, without loss of generality. The particle's specific energy $E$ and specific angular momentum $L$ are given by
\begin{align}
	 E &= f(r_p)u^t\label{eq:tdot},\\
	 L &= r_p^2 u^\varphi\label{eq:phidot},
\end{align}
and both are conserved along the orbit. When Eqs.~\eqref{eq:tdot} and \eqref{eq:phidot} are combined with the timelike normalization condition $u^{\alpha}u_{\alpha} = -1$, we obtain the radial equation of motion,
\begin{align}
	(u^r)^2 = E^2 - V(r_p; L) \label{eq:RadialEqn},
\end{align}
where the effective potential is
\begin{align}
	V(r; L) := f(r)\left(1+\frac{L^2}{r^2}\right).
\end{align}

For a hyperbolic scattering orbit we require $r_p \rightarrow \infty$ with $|u^r| \rightarrow u^r_{\infty} > 0$ as $\tau \rightarrow \pm\infty$, which from Eq.~\eqref{eq:RadialEqn} is possible if, and only if, 
\begin{align}
    E > 1. \label{eq:E_condition}
\end{align}
For the particle to scatter back to infinity (rather than be captured by the central object), we additionally require that 
\begin{align}
    %L &> \frac{M}{\sqrt{{E^2-1}}}\sqrt{(27E^4+9\nu E^3 - 36E^2 - 8\nu E+8)/2} , \label{eq:L_condition}
    L &> M\sqrt{\frac{27E^4+9\nu E^3 - 36E^2 - 8\nu E+8}{2(E^2-1)}} , \label{eq:L_condition}
\end{align}
where $\nu := \sqrt{9E^2 - 8}$ \cite{LongBarack2021}. Note that we may take $L > 0$ without loss of generality due to reflectional symmetry in the equatorial plane. Provided conditions \eqref{eq:E_condition} and \eqref{eq:L_condition} are both satisfied, the right-hand side of Eq.~\eqref{eq:RadialEqn} has three real roots $r_1, r_2$, and $\rmin$, with $ r_1 < 0$ and $2M < r_2 < r_{\rm min}$ (explicit expressions for these roots are given in Eqs.~(12)--(14) of Ref.~\cite{Whittall:2023xjp}). The radial motion thus begins at $r_p \rightarrow \infty$, approaches the turning point at $r_p = \rmin$ (the {\it periapsis radius}), and then returns to $r_p \rightarrow \infty$. 

\begin{figure}[tb]
  \centering
  \includegraphics[width=\linewidth]{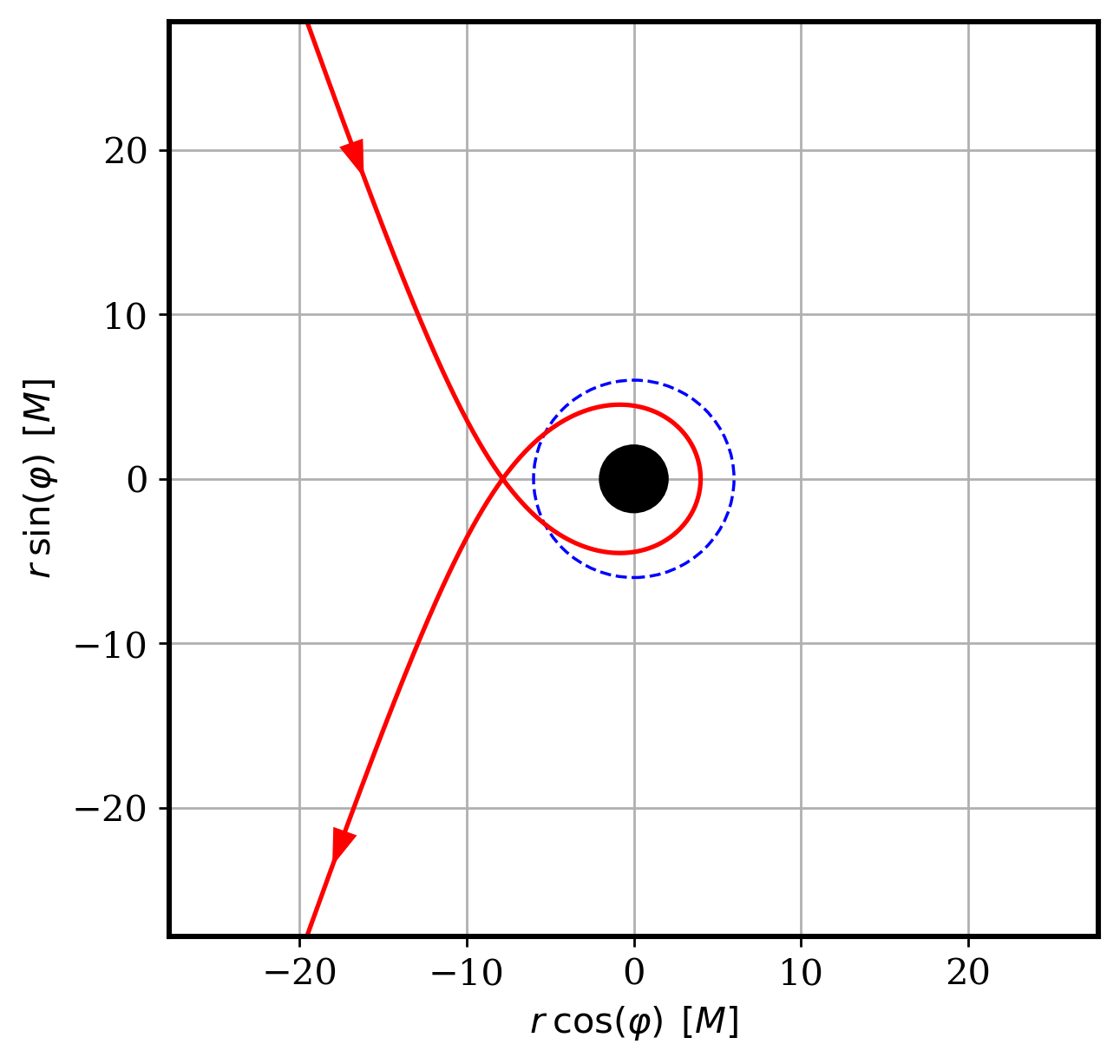}~\\~\\
  \caption{\label{fig:E1pt1_rmin4M_geodesic}Illustration of the reference scattering geodesic orbit with parameters $E = 1.1$ and $\rmin = 4M$. The central black hole (black disk) and its innermost stable circular orbit (blue circle) are drawn to scale.}
\end{figure}

A scattering geodesic is uniquely specified by the pair $(E, L)$, or by the pair $(E, \rmin)$. The latter set of parameters can be obtained from the former using the expression for $\rmin(E, L)$ from Eq.~(14) of Ref.~\cite{Whittall:2023xjp}, while the reverse transformation is most easily achieved by rearranging Eq.~\eqref{eq:RadialEqn} (with $u^r = 0$) into an expression for $L(E, \rmin)$. Figure~\ref{fig:E1pt1_rmin4M_geodesic} displays an example scattering geodesic with energy $E = 1.1$ and periapsis radius $\rmin = 4M$ (exactly), corresponding to angular momentum $L \approx 4.767M$. This will be our standard reference geodesic from this point onward, with all calculations performed for a scalar charge moving along this orbit.

\subsection{Scalar-field self-force}
The particle sources a scalar field $\Phi$ (assumed to be massless and minimally coupled) which obeys the Klein-Gordon equation on the background Schwarzschild spacetime, 
\begin{align}
	\nabla_{\mu}\nabla^{\mu}\Phi = -4\pi T. \label{eq:ScalarEOM}
\end{align}
Here $\nabla_\mu$ is the covariant derivative compatible with the background Schwarzschild geometry, and $T$ is the scalar charge density,
\begin{align}
	T\left(x^{\alpha}\right) := q\displaystyle\int_{-\infty}^{+\infty} \delta^{4}\left(x^{\alpha} - x_p^{\alpha}(\tau)\right)\frac{d\tau}{\sqrt{-g(x)}}, \label{eq:ScalarSource}
\end{align}
with $g$ being the determinant of the Schwarzschild metric. 

Backreaction from the scalar field produces a small acceleration---the ``self-force'' effect---that drives the worldline away from the background geodesic,
\begin{align}
    u^{\beta}\nabla_{\beta}\left(\mu u^\alpha\right) = F_{\text{self}}^{\alpha} \propto q^2. \label{eq:scalar_self_forced_EOM}
\end{align}
Note that we neglect for simplicity the gravitational self-force acting on the particle, and the effect of the scalar field's backreaction on the background spacetime itself. 

A standard way to calculate the self-force in Eq.~\eqref{eq:scalar_self_forced_EOM} is the mode-sum regularization procedure introduced in Refs.~\cite{Barack:1999wf, Barack01}. In this approach, the retarded field is first decomposed into spherical harmonics defined on spheres of constant Schwarzschild radius $r$, 
\begin{align}
	\Phi &= \sum_{\ell m} \frac{1}{r}\psi_{\lm}(t, r) Y_{\ell m}(\theta, \varphi)\label{eq:spherharmPhi}.
\end{align}
The self-force at proper time $\tau$ along the orbit may be computed from the retarded field modes $\psi_{\lm}(t,r)$ using the {\it mode-sum formula},
\begin{align}
    F_{\alpha}^{\text{self}}(\tau) = \sum_{\ell=0}^{\infty}\SB{F_{\alpha\pm}^{(\text{full})\ell}(\tau) - \SP{\ell+\frac{1}{2}}A_{\alpha}^{\pm}(\tau) - B_{\alpha}(\tau)},\label{eq:mode_sum_formula}
\end{align}
where 
\begin{align}
    F_{\alpha\pm}^{(\text{full})\ell}(\tau) := \lim_{x \rightarrow x_p^{\pm}(\tau)} q\sum_{m=-\ell}^{\ell}\nabla_{\alpha}\SB{\frac{1}{r} \psi_{\lm}(t,r)Y_{\lm}(\theta,\vf)},
\end{align}
with the $\pm$ denoting whether the limit is taken from $r \rightarrow r_p(\tau)^-$ or $r\rightarrow r_p(\tau)^+$. Note that $F_{\alpha\pm}^{(\text{full})\ell}$ is finite for each $\ell$. The quantities $A^{\pm}_{\alpha}$ and $B_{\alpha}$ are known as the {\it regularization parameters}, first derived analytically for generic Schwarzschild geodesics in Refs.~\cite{regpar, Barack:2001gx}. 

The mode-sum regularization approach has been widely used for both scalar-field and first-order gravitational self-force calculations along bound orbits (including the state-of-the-art code for inclined-eccentric Kerr geodesics described in Ref.~\cite{vandeMeent2018}), and has been the chosen approach for all numerical calculations of the (scalar-field) self-force along scattering orbits that have been performed thus far \cite{BarackLong2022, Whittall:2023xjp}. 

\section{Frequency-domain self-force calculations}\label{sec:FDmethods}
The primary numerical inputs to the mode-sum formula \eqref{eq:mode_sum_formula} are the time-domain scalar-field modes $\psi_{\lm}(t,r)$ and their derivatives. In this section we review how these quantities are obtained by decomposing them into frequency modes. The scalar-field equation~\eqref{eq:ScalarEOM} is then reduced to a radial ordinary differential equation, whose inhomogeneous solution can be obtained using the standard method of variation of parameters. Section~\ref{sec:direct_reconstruction} then illustrates how the Gibbs phenomenon renders impractical a naive reconstruction of the time-domain modes from these inhomogeneous frequency modes. Section~\ref{sec:EHS_reconstruction} introduces the EHS method as a way to avoid the Gibbs phenomenon, and discusses the limitations of this approach for scattering and highly eccentric bound orbits.

\subsection{Frequency-domain decomposition}\label{sec:FD_decomp}
Following the spherical harmonic decomposition of Eq.~\eqref{eq:spherharmPhi}, Eq.~\eqref{eq:ScalarEOM} becomes
\begin{align}
	-\frac{\partial^2\psi_{\lm}}{\partial t^2}+\frac{\partial^2\psi_{\lm}}{\partial r_{*}^2} - V_\ell(r)\psi_{\lm}= -4\pi rf(r) T_{\lm}, \label{eq:ScalarEOMTD}
\end{align}
where $r_* := r + 2M\log\SP{\frac{r}{2M}-1}$ is the Regge-Wheeler tortoise coordinate, the potential $V_{\ell}(r)$ is 
\begin{align}
    V_{\ell}(r) := \left(\frac{\ell(\ell+1)}{r^2}+\frac{2M}{r^3}\right)f(r) \label{eq:ScalarEffPot},
\end{align}
and 
\begin{align}
    T_{\lm}(t,r) :=& \displaystyle\int d^2\Omega \>Y^{*}_{\lm}(\theta,\varphi) T(t,r,\theta,\vf)\\
                =&\: \frac{q Y_{\lm}^*\left(\frac{\pi}{2},0\right)}{r_p^2(t)u^t(t)}e^{-im\varphi_p(t)}\delta(r-r_p(t))\label{eq:TDsource}
\end{align}
are the spherical harmonic modes of the scalar charge density.

Equation~\eqref{eq:ScalarEOMTD} may be solved numerically as a partial differential equation to obtain the time-domain modes $\psi_{\lm}(t,r)$, or---as we shall do here---by additionally separating it into frequency modes:
\begin{align}
	\psi_{\lm}(t,r) &= \displaystyle\int_{-\infty}^{+\infty}d\omega\, e^{-i\omega t}\psi_{\lm\omega}(r) \label{eq:freq_decomp_field}.
\end{align} 
Equation \eqref{eq:ScalarEOMTD} then reduces to an ordinary differential equation for each $\omega$,
\begin{align}
	\frac{d^2\psi_{\lm\omega}}{dr_{*}^2} - \left(V_\ell(r)-\omega^2\right)\psi_{\lm\omega} = S_{\lmw}(r), \label{eq:ScalarEOMFD}
\end{align}
where 
\begin{align}
    S_{\lmw}(r) := -4\pi r f(r) T_{\lmw}(r), \label{eq:Slmw_def}
\end{align}
with
\begin{align}
    T_{\lmw}(r) &:= \frac{1}{2\pi}\int_{-\infty}^{+\infty} dt \>e^{i\omega t}T_{\lm}(t, r)\nonumber\\&= \frac{qY_{\lm}^*\left(\frac{\pi}{2},0\right)}{\pi r^2 |u^r(r)|}\cos\left(\omega t_p(r) - m\varphi_p(r)\right)\Theta(r-r_{\rm min}).\label{eq:SourceFT}
\end{align}
The relations $u^r(r)$, $t_p(r)$ and $\varphi_p(r)$ are evaluated on the outbound, $u^r > 0$, leg of the scattering geodesic.

The frequency-domain equation~\eqref{eq:ScalarEOMFD} can be solved using the method of variation of parameters, as outlined in Secs. II D and II E of Ref.~\cite{Whittall:2023xjp}. The first step in this process is to introduce a basis of solutions, $\psi_{\ell\omega}^\pm$, each of which obeys the homogeneous ($S_{\lmw}=0$) version of Eq.~\eqref{eq:ScalarEOMFD} subject to the physical, retarded boundary conditions at precisely {\it one} of the boundaries $r_* \rightarrow  +\infty$ or $r_* \rightarrow -\infty$. In particular, for $\omega \neq 0$, we choose boundary conditions
\begin{align}
    \psi_{\lw}^-(r) &\sim e^{-i\omega r_{*}}\>\>\text{  as } r_{*} \rightarrow -\infty \label{eq:BChorizon},\\
    \psi_{\lw}^+(r) &\sim e^{i\omega r_{*}} \>\>\>\> \text{  as } r_{*}\rightarrow +\infty  \label{eq:BCinfty},
\end{align}
corresponding to radiation which is purely outgoing at future null infinity ($\psi_{\lw}^+$), and purely ingoing at the future event horizon ($\psi_{\lw}^-$). The static ($\omega=0$) homogeneous solutions are given analytically by
\begin{align}
    \psi_{\ell0}^-(r) &:= r P_{\ell}\SP{\frac{r-M}{M}},\label{eq:psi_minus_def} \\
    \psi_{\ell0}^+(r) &:= r Q_{\ell}\SP{\frac{r-M}{M}},\label{eq:psi_plus_def]}
\end{align}
where $P_\ell$ and $Q_\ell$ are the Legendre polynomial and Legendre function of the second kind, respectively. These solutions are regular at the horizon and at infinity, respectively.

The inhomogeneous solution to Eq.~\eqref{eq:ScalarEOMFD} obeying retarded boundary conditions at {\it both} boundaries can then be expressed in general as
\begin{align}
    	\psi_{\lm\omega}(r) =  \>&\psi_{\lw}^+(r)\displaystyle\int_{r_{\rm min}}^r\frac{\psi_{\lw}^-(r')S_{\lm\omega}(r')}{W_{\lw}}\frac{dr'}{f(r')}\label{eq:VoPRadialEqnScattering} \\\nonumber &+ \psi_{\lw}^-(r)\displaystyle\int_{r}^{+\infty}\frac{\psi_{\lw}^+(r')S_{\lm\omega}(r')}{W_{\lw}}\frac{dr'}{f(r')}, 
\end{align}
where $W_{\lw}:=\psi_{\lw}^-\frac{d\psi_{\lw}^+}{dr_*}- \psi_{\lw}^+\frac{d\psi_{\lw}^-}{dr_*}$ is the Wronskian of the homogeneous solutions, which depends only on $\ell$ and $\omega$, and not on $r$. The sole exception to Eq.~\eqref{eq:VoPRadialEqnScattering} is the case of the static monopole, $\omega = 0 = \ell$, for which the integral in the second line of Eq.~\eqref{eq:VoPRadialEqnScattering} diverges logarithmically at the upper limit. In fact, the true retarded solution must also diverge like $\psi_{000} \sim r \log r$ as $r\rightarrow \infty$. The time-domain solution, however, does not diverge, and this technicality will not affect our numerical tests in Sec. V.

\subsection{Direct reconstruction and the Gibbs phenomenon}\label{sec:direct_reconstruction}

\begin{figure*}[tbh!]
  \centering
  \includegraphics[width=\linewidth]{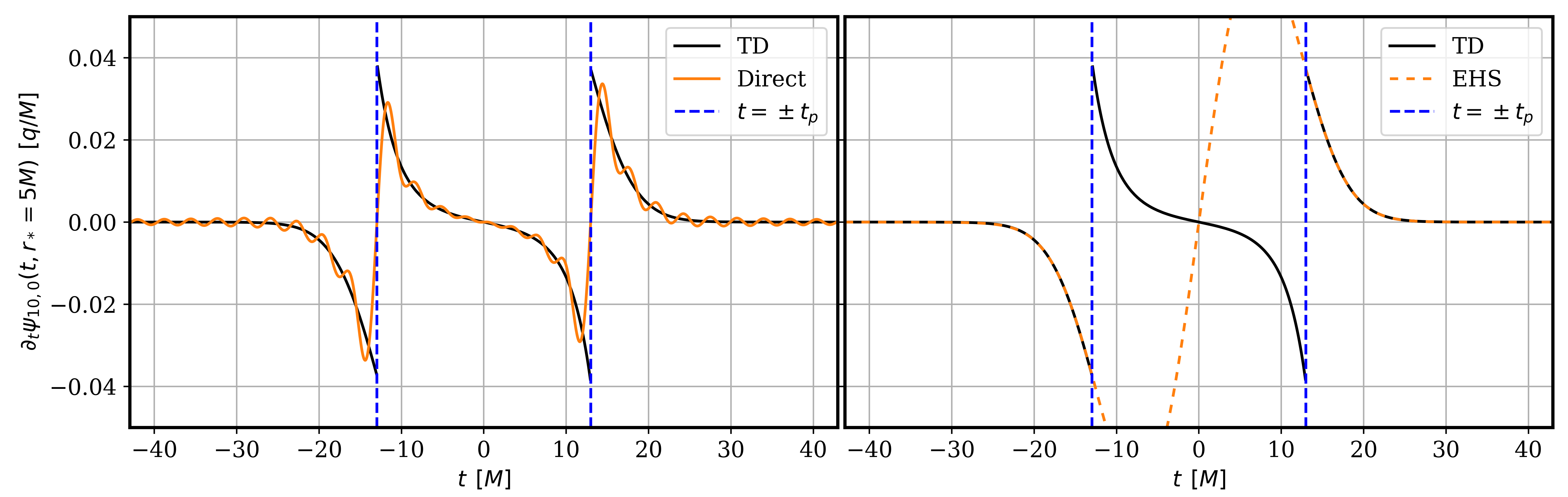}~\\~\\
  \caption{\label{fig:direct_vs_ehs_vs_td}Value of $\partial_t\psi_{\lm}(t,r)$ for $\ell=10$, $m=0$ as a function of time along the spatial slice $r_* = 5M$ as reconstructed directly from the retarded field $\psi_{\lmw}(r)$ (left panel, orange) and from the internal EHS $\tilde\psi^-_{\lmw}(r)$ (right panel, orange, dashed), both using frequency truncation $\wmax= 1.954/M$. The ``correct'' result, obtained using the time-domain code of Ref.~\cite{BarackLong2022} (black), is included for comparison. {\it Left panel:} the direct curve tracks the time-domain curve throughout, but is severely afflicted by Gibbs ringing, especially in the vicinity of the discontinuities at the particle crossing times, $t = \pm t_p(r_* = 5M)$ (indicated with vertical blue lines), and converges to the two-sided average at those times. {\it Right panel:} the internal EHS curve coincides with the time-domain result for $|t| > t_p$, with the correct one-sided limits as $t\to \pm t_p^\pm$ , but disagrees completely in $|t| < t_p$, as expected. }
\end{figure*}

The presence of a delta function on the right-hand side of Eq.~\eqref{eq:ScalarEOMTD} [recall Eq.~\eqref{eq:TDsource}] means that, in general, the derivatives of $\psi_{\lm}(t,r)$ are discontinuous at the particle's worldline, $r = r_p(t)$. This has important consequences for frequency-domain self-force calculations, rendering impractical a naive reconstruction of the time-domain field $\psi_{\lm}(t,r)$ and their derivatives from the inhomogeneous frequency modes given in Eq.~\eqref{eq:VoPRadialEqnScattering}. The problem is the well-known Gibbs phenomenon encountered when attempting to represent discontinuous functions globally with a series of smooth basis functions. In particular, the Fourier series/integrals for the derivatives converge toward the correct values everywhere off the worldline, but only very slowly (with terms decaying like $1/\omega$), conditionally, and nonuniformly near to the worldline. On the worldline itself the series is expected to converge to the two-sided average of the derivative. For the modes $\psi_{\lm}(t,r)$ themselves (which are continuous), the series converges to the correct value everywhere, but with terms which decay only slightly faster, like $1/\omega^2$. 

A numerical illustration of the Gibbs phenomenon is given in the left panel of Fig.~\ref{fig:direct_vs_ehs_vs_td}, which shows $\partial_t\psi_{\lm}$ for the mode $(\ell, m) = (10, 0)$. The derivative is plotted as a function of time along the spatial slice $r_* = 5M$, as calculated (a) using the fiducial time-domain numerical code from Ref.~\cite{BarackLong2022} (hereafter referred to simply as the TD code) and (b) by reconstruction from the inhomogeneous frequency modes given in Eq.~\eqref{eq:VoPRadialEqnScattering}, including all modes with frequency $|\omega| < 1.954/M =: \wmax$ in the Fourier integral,
\begin{align}
    \partial_t \psi_{\lm}(t,r) \approx - \displaystyle\int_{-\wmax}^{+\wmax} d\omega\>e^{-i\omega t}\>i\omega\>\psi_{\lmw}(r).\label{eq:partial_fourier_tderiv}
\end{align}
For further details of the numerical calculation, see Appendix~\ref{app:numerical_methods}. The derivative is discontinuous at $t = \pm t_p(r_* = 5M) \approx \pm 13.0M$, corresponding to the times when the particle's worldline intersects the sphere $r_* = 5M$. The value reconstructed from the truncated frequency modes displays characteristic Gibbs ``ringing", oscillating around the true value with increasing amplitude in the vicinity of the discontinuities. At the points $t = \pm t_p$ themselves, the reconstructed curve can be seen to pass through the midpoint of the two 1-sided limits of the true result, as expected.

In this paper we will use the term {\it direct reconstruction} to refer to the naive reconstruction of a time-domain field from its Fourier coefficients using partial Fourier integrals or series in the standard way, as in Eq.~\eqref{eq:partial_fourier_tderiv}. This approach is generally insufficient in practice (i.e, with practical values of the frequency truncation $\wmax$) for reconstructing the time-domain modes of the retarded field $\psi_{\lm}(t,r)$ and its derivatives to the precision required for the self-force mode sum, Eq.~\eqref{eq:mode_sum_formula}. For this task, alternative techniques are required.

\subsection{Reconstruction from the EHS and its limitations}\label{sec:EHS_reconstruction}
The method of EHSs was introduced in Ref.~\cite{Barack:2008ms} to circumvent the Gibbs phenomenon in the case of frequency-domain self-force calculations along eccentric bound orbits. Rather than attempting to reconstruct the nonsmooth field $\psi_{\lm}(t,r)$ globally from a single set of frequency modes, this approach instead constructs {\it analytic} (smooth) functions that agree with the physical time-domain modes on only one side of the discontinuity, i.e., in either $r < r_p(t)$ or $r > r_p(t)$. By working with analytic functions in this way, one only has to evaluate Fourier series or integrals with exponential, uniform convergence. 

To illustrate the method, we show how EHSs can be used to reconstruct $\psi_{\lm}(t,r)$ in the {\it internal region} $r \leq r_p(t)$ in the case of a scattering orbit, as presented in Ref.~\cite{Whittall:2023xjp}, based on the original bound-orbit calculation from Ref.~\cite{Barack:2008ms}. The first step is to note that the frequency-domain internal EHS, defined by
\begin{align}
    \tilde\psi_{\lmw}^-(r) := C_{\lmw}^-\psi_{\lw}^-(r), \label{eq:int_ehs_def}
\end{align}
with the normalization integral
\begin{align}
    C_{\lmw}^- := \displaystyle\int_{\rmin}^{+\infty}\frac{\psi_{\lw}^+(r')S_{\lm\omega}(r')}{W_{\lw}}\frac{dr'}{f(r')}, \label{eq:Cminus_def}
\end{align}
coincides exactly with the inhomogeneous frequency modes \eqref{eq:VoPRadialEqnScattering}, by construction, for $r \leq \rmin$. Defining the time-domain internal EHS,
\begin{align}
    \tilde\psi_{\lm}^-(t,r) := \displaystyle\int_{-\infty}^{+\infty} d\omega\>e^{-i\omega t}\>\tilde\psi_{\lmw}^-(r),\label{eq:intEHS_TD}
\end{align}
we thus have
\begin{align}
    \psi_{\lm}(t,r) = \tilde\psi_{\lm}^{-}(t,r) \quad \forall\,  t,\, r  \leq \rmin. \label{eq:vacuum_equality}
\end{align}
The remainder of the argument relies on analytic continuation. It was demonstrated analytically in Ref.~\cite{Barack:2008ms} that $C_{\lmw}^-$ decays exponentially with $\omega$ for bound orbits, and this was confirmed numerically for the scattering case also in Ref.~\cite{Whittall:2023xjp}. This ensures that $\tilde\psi_{\lm}^-(t,r)$ is analytic in $t$ and $r$ everywhere. The physical retarded field $\psi_{\lm}^-(t,r)$ is likewise expected to be analytic throughout $r < r_p(t)$, or else irregularities would have to arise in the vacuum region. It follows that Eq.~\eqref{eq:vacuum_equality} can be extended to all of $r \leq r_p(t)$,
\begin{align}
    \psi_{\lm}(t,r) = \tilde\psi_{\lm}^{-}(t,r) \quad \forall  \leq r_p(t). \label{eq:LHS_equality}
\end{align}

\begin{figure}[t]
  \centering
  \includegraphics[width=\linewidth]{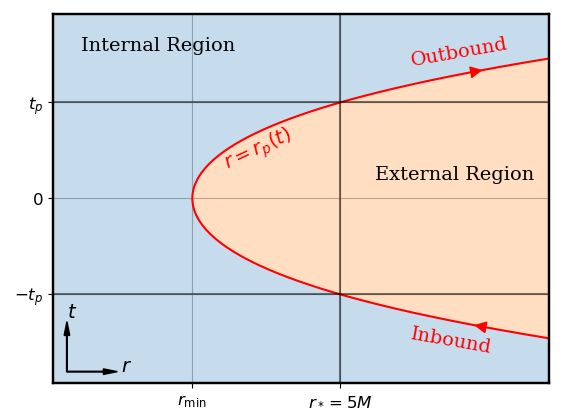}~\\~\\
  \caption{\label{fig:int_ext_regions}Illustration of the {\it internal} $[r < r_p(t)]$ and {\it external} $[r > r_p(t)]$ regions of the $(t,r)$ plane. For a fixed radius with $r_* = 5M$, the point $(t,r)$ lies in the external region whenever $|t| \leq t_p(r_*=5M)$, and in the internal region when $t > t_p$ or $t < -t_p$. Here $\pm t_p(r_*=5M) \approx \pm 13.0M$ are the times at which the particle crosses the sphere $r_*=5M$.}
\end{figure}

The effectiveness of this approach is illustrated in the right panel of Fig.~\ref{fig:direct_vs_ehs_vs_td}, which again displays $\partial_t\psi_{\lm}$ as a function of time at fixed $r_* = 5M$ for the $(\ell, m) = (10, 0)$ mode, comparing the internal EHS reconstruction to a reference result from the TD code. The field reconstructed from the internal EHS modes is observed to coincide with the reference result in the disjoint regions $t < -t_p(r_*=5M) \approx -13.0M$ and $t > t_p(r_*=5M)$, corresponding to the times when $r_p^*(t) > 5M$, i.e.\ when the field point $(t, r_*=5M)$ lies in the internal region (see Fig.~\ref{fig:int_ext_regions}). This is precisely the region where Eq.~\eqref{eq:LHS_equality} holds. During the times $|t| < t_p(r_*=5M)$, however, the internal EHS field disagrees completely with the reference solution; the internal EHS cannot be used to reconstruct the physical solution in the external region $r > r_p(t)$.

For bound-orbit sources, an equivalent construction can be used to obtain the field in the external region by choosing the homogeneous solution $\tilde\psi^+_{\lmw}(r)$ that coincides with the physical frequency modes in the vacuum region beyond the apoapsis radius, $r > \rmax$. This is not possible in scattering scenarios (for which the motion stretches all the way to radial infinity), and currently no alternative EHS reconstruction approach has been demonstrated for the external region outside a scattering orbit. This is not a critical issue for the scalar-field self-force model, for which it is straightforward to calculate the self-force from the internal EHS field using one-sided regularization, i.e., using only the $r \rightarrow r_p(t)^-$ limit in the mode sum, Eq.~\eqref{eq:mode_sum_formula}. The restriction is more problematic, however, when extending to the full gravitational self-force problem, for which the conventional approach is to calculate the self-force from modes of the metric perturbation in a so-called radiation gauge, using a two-sided mode-sum regularization approach \cite{Pound:2013faa}. A one-sided regularization procedure in the radiation gauge has also been formulated \cite{Pound:2013faa} but it is much more complicated, and its applicability is yet to be demonstrated in practice. 

The precision of EHS reconstruction is also sorely tested by the occurrence of severe numerical cancellation, as first noted in Ref.~\cite{vandeMeent2016}. The origin of this problem lies in the unphysical behavior of the EHS frequency modes in the source region. Considering the internal EHS, for example, the low-frequency EHS modes $\tilde\psi^-_{\lmw}(r)$ behave like the static modes in Eq.~\eqref{eq:psi_minus_def}, growing polynomially like $\tilde\psi_{\lmw}^-(r) \sim r^{\ell+1}$. The physical inhomogeneous modes $\psi_{\lmw}$, on the other hand, do not exhibit this growth. The agreement in Eq.~\eqref{eq:LHS_equality} thus requires a growing level of cancellation between the low-frequency EHS modes in Eq.~\eqref{eq:intEHS_TD} as $r$ is increased above $\rmin$, with a loss of precision that grows exponentially with $\ell$ (see, e.g., Fig.~2 of Ref.~\cite{vandeMeent2016} and Fig.~11 of Ref.~\cite{Whittall:2023xjp}). For a bound orbit, the greatest cancellation in $\tilde\psi^-_{\lm}(t,r)$ occurs at the apoapsis $\rmax$, and the loss of precision at this point will be greater when $\rmax/\rmin$ is greater, so that the impact of the cancellation problem grows with orbital eccentricity. This situation reaches its climax in the case of scattering, for which there is no upper limit on the orbital radius, and the degree of cancellation encountered increases without bound as one attempts to reconstruct the time-domain field at ever larger radii. 

Despite the cancellation problem, significant success has still been achieved using EHS reconstruction for self-force calculations. In the case of bound orbits, Refs.~\cite{vandeMeent2016} and \cite{vandeMeent2018} computed the self-force along generic (i.e.~eccentric and inclined) bound geodesics in the Kerr spacetime, using high-precision arithmetic to compute all frequency-domain quantities to sufficiently high accuracy as to retain sufficient precision after cancellation, albeit at great computational cost. This approach was deemed impractical for scattering calculations in Ref.~\cite{Whittall:2023xjp}, owing to the significantly greater computational cost already inherent to scattering calculations, and to the presence of additional sources of error in the calculation of the frequency modes which could not be easily reduced even with high-precision arithmetic. Instead, Ref.~\cite{Whittall:2023xjp} introduced a scheme which dynamically selected the upper truncation $\ell_{\rm max}$ for the mode sum \eqref{eq:mode_sum_formula} so as to detect and exclude problematic high-cancellation large-$\ell$ modes. However, the cost of this approach is a rapid loss of precision as $r$ increases due to the neglected large-$\ell$ tail, which significantly limits the radius out to which we may accurately compute the self-force, and in turn prevents us from accurately calculating important observables such as the scattering angle \cite{Whittall:2023xjp}. 

\section{The Gegenbauer reconstruction procedure}\label{sec:Geg_intro}
We have seen that direct reconstruction from the inhomogeneous modes $\psi_{\lmw}$ is impractical due to the Gibbs phenomenon, while reconstruction from the unphysical frequency-domain EHS modes suffers from catastrophic cancellation in the high-eccentricity limit. The purpose of this section is to introduce a third reconstruction approach, never before applied to the self-force problem, through which the retarded time-domain field modes $\psi_{\lm}(t,r)$ may be obtained with exponential accuracy from the physical inhomogeneous modes $\psi_{\lmw}$.

The foundation of this new approach is the concept of a {\it Gibbs complement} to the Fourier basis in which we expand the scalar field. The Gibbs complement is a new complete basis of functions (defined on a given compact time interval) that obeys two important conditions \cite{gottlieb1998general}: (1) for a function that is analytic on the compact time interval, the expansion in the complementary basis should converge exponentially, and (2) the projection of the high-frequency Fourier modes onto the low-degree modes of the complementary basis can be made exponentially small. The precise meaning of these conditions will be detailed in Sec.~\ref{sec:geg_process}, where we describe how a piecewise analytic function may be reconstructed on a domain on which it is analytic by reprojecting its partial Fourier representation onto a Gibbs complementary basis of {\it Gegenbauer polynomials}, as originally developed in Refs.~\cite{gottlieb1992gibbs, gottlieb1994resolution, gottlieb1995gibbs, gottlieb1997gibbs}. This Gegenbauer reconstruction approach is then illustrated numerically for an analytical continuous-spectrum  example in Sec.~\ref{sec:geg_toy_example}.

\subsection{Process}\label{sec:geg_process}
Let $f(s)$ be a piecewise analytic function with Fourier transform $\hat f(\omega)$, related by
\begin{align}
    f(s) = \displaystyle\int_{-\infty}^{+\infty}\hat f(\omega) e^{-i\omega s}d\omega \label{eq:toy_f}.
\end{align}
The partial Fourier integrals with truncation frequency $\wmax > 0$ are defined as
\begin{align}
    F(s; \wmax) := \displaystyle\int_{-\wmax}^{+\wmax} \hat f(\omega) e^{-i\omega s} d\omega. \label{eq:F_partial_fourier}
\end{align}

We assume without loss of generality that $f(s)$ is analytic on the interval $s \in [-1,1]$. In this case, we may also expand $f(s)$ in a basis of {\it Gegenbauer polynomials} $C_k^\lambda(s)$ on that interval,
\begin{align}
    f(s) = \sum_{k=0}^{\infty} f_k^{\lambda}C_k^{\lambda}(s) \qquad \left(s \in [-1,1]\right), \label{eq:f_geg_series}
\end{align}
and this series converges exponentially at fixed $\lambda > -1/2$. 

The partial Gegenbauer sums are denoted as
\begin{align}
    \mathcal{F}_N(s) := \sum_{k=0}^N f_k^\lambda C_k^\lambda(s),
\end{align}
where the dependence of $\mathcal{F}_N$ on $\lambda$ is kept implicit.

The properties of the Gegenbauer polynomials are discussed in greater detail in Appendix~\ref{app:gegenbauer_polys}, but here it suffices to note that $C_k^\lambda$ is a polynomial of degree $k$ for each $k$, and they are collectively orthogonal on $[-1,1]$ with respect to the weight function $(1-s^2)^{\lambda-1/2}$,
\begin{align}
    \int_{-1}^{1}(1-s^2)^{\lambda - 1/2}C_n^\lambda(s)C_m^\lambda(s) ds = h_n^\lambda \delta_{nm},\label{eq:geg_inner_product}
\end{align}
where closed-form expressions for $h_n^\lambda$ are given in Appendix~\ref{app:gegenbauer_polys}. The Gegenbauer coefficients $f_k^{\lambda}$ may hence be obtained via 
\begin{align}
    f_k^{\lambda} = \frac{1}{h_k^\lambda} \int_{-1}^{1} (1-s^2)^{\lambda-1/2}f(s)C_k^\lambda(s) ds .
\end{align}

The partial Fourier integrals $F(s;\wmax)$ may also be expanded on $-1 \leq s \leq 1$ for any $\wmax$,
\begin{align}
    F(s;\wmax) = \displaystyle\sum_{k=0}^{\infty} g_k^{\lambda}(\wmax) C_k^\lambda(s),
\end{align}
with
\begin{align}
    g_k^{\lambda}(\wmax) = \frac{1}{h_k^\lambda} \int_{-1}^{1} (1-s^2)^{\lambda-1/2}F(s; \wmax)C_k^\lambda(s) ds. \label{eq:g_k_def}
\end{align}
The corresponding partial Gegenbauer series is
\begin{align}
    \mathcal{G}_N(s) := \sum_{k=0}^N g_k^\lambda(\wmax) C_k^\lambda(s),\label{eq:G_N_def}
\end{align}
which implicitly depends on $\lambda$ and $\wmax$. In the Gegenbauer reconstruction method, we will use $\mathcal{G}_N(s)$ as an approximant for the original function $f(s)$. Using the triangle inequality, the error of this approximation obeys 
\begin{align}
    \|f - \mathcal{G}_N \|_{\infty} \leq \|f - \mathcal{F}_N \|_{\infty} + \|\mathcal{F}_N - \mathcal{G}_N \|_{\infty}, \label{eq:err_bound}
\end{align}
where
\begin{align}
    \|h\|_{\infty} := \sup_{-1 \leq s \leq 1} \Big|h(s)\Big|
\end{align}
is the $L_{\infty}$ norm on $-1 \leq s \leq 1$. The first term on the right-hand side of Eq.~\eqref{eq:err_bound} is referred to as the \textit{regularization error}, committed by expanding $f(t)$ as a Gegenbauer series on $[-1, 1]$. The second is the \textit{truncation error} in this series arising from the truncation of the Fourier integral at finite frequency.  Note that
\begin{align}
    \mathcal{F}_N(s) - \mathcal{G}_N(s) &= \sum_{k=0}^N \left(f_k^\lambda - g_k^\lambda\right)C_k^{\lambda}(s)\\
    &= \sum_{k=0}^N  \bar g_k^\lambda C_k^\lambda(s),
\end{align}
where 
\begin{align}
    \bar g_k^\lambda = \frac{1}{h_k^\lambda}\displaystyle\int_{-1}^{+1}(1-s^2)^{\lambda-1/2} \bar F(s;\wmax) C_k^\lambda(s),
\end{align}
and
\begin{align}
    \bar F(s; \wmax) := f(s) - F(s;\wmax)
\end{align}
is the $\omega > \wmax$ Fourier content of the function $f(s)$. $\mathcal{F}_N - \mathcal{G}_N$ is thus the projection of the high-frequency Fourier modes of $f(s)$
onto the span of the low-degree Gegenbauer polynomials. 

The defining properties (1) and (2) of the Gibbs complementary basis are the requirements that the regularization and truncation errors can both be made to decay exponentially by taking $\wmax$ and $N$ to infinity. It then follows from Eq.~\eqref{eq:err_bound} that approximant \eqref{eq:G_N_def} converges uniformly and exponentially to the true function $f(s)$ on $[-1,1]$. Remarkably, it has been shown that both errors do indeed decay exponentially if $N$ {\it and} $\lambda$ are taken to infinity in linear proportion to $\wmax$ \cite{gottlieb1992gibbs, gottlieb1994resolution, gottlieb1995gibbs, gottlieb1997gibbs}, i.e.
\begin{align}
    N = \lfloor \alpha\wmax \rfloor, \quad \lambda = \beta\wmax, \label{eq:param_prop}
\end{align}
for some constants $\alpha > 0$ and $\beta > 0$. The possible values of the proportionality constants that ensure exponential convergence are not unique, but the choice can affect the rate of convergence and accuracy of the resulting approximant. We note that the exponential convergence was originally established only for the discrete-spectrum case, i.e., when the continuous Fourier decomposition in Eq.~\eqref{eq:toy_f} is replaced by a discrete Fourier series on a compact interval. Nonetheless, intuition (validated by the numerical results of the following sections) strongly supports the applicability of the Gegenbauer reconstruction approach to the continuous frequency problem. We will now demonstrate this with a toy analytical example.

\subsection{Example}\label{sec:geg_toy_example}
\begin{figure*}[tbh!]
  \centering
  \includegraphics[width=\textwidth]{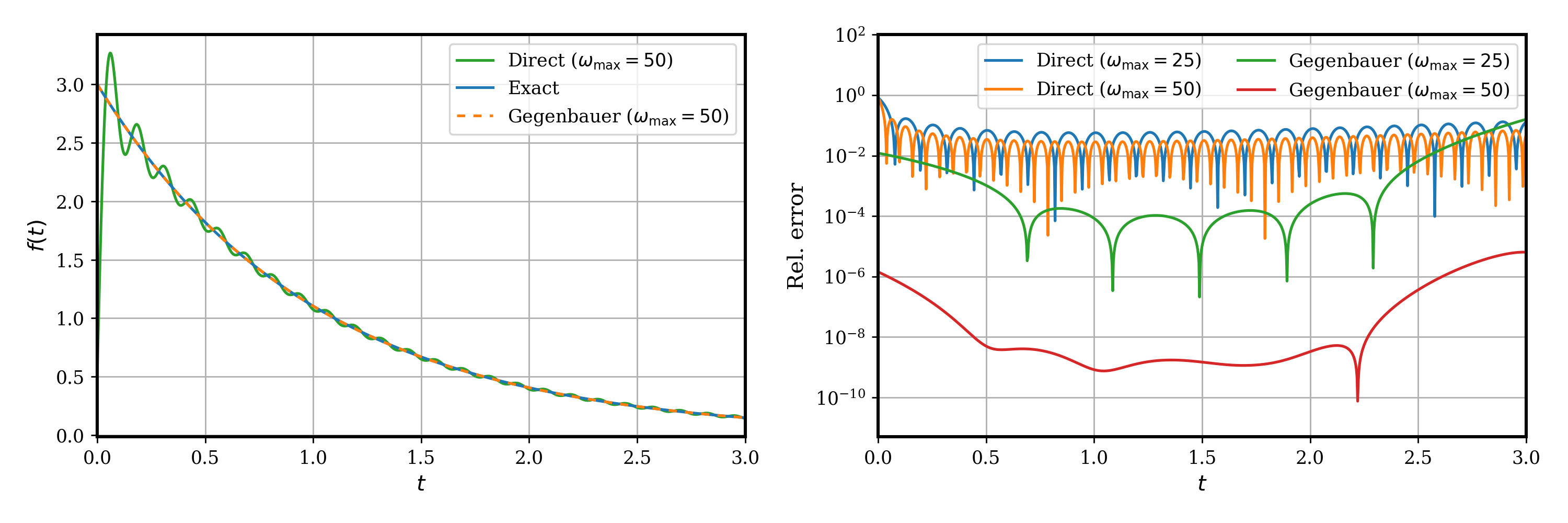}~\\~\\
  \caption{\label{fig:toy_example}{\it Left panel:} the exact function $f(t)$ from Eq.~\eqref{eq:example_f} on $0 \leq t \leq 3$ (blue), and its direct [Eq.~\eqref{eq:F_partial_fourier}, green] and Gegenbauer [Eq.~\eqref{eq:geg_approximant_example}, orange dashed] reconstructions, both using a frequency truncation $\omega_{\rm max} = 50$ and parameters $(\alpha, \beta) = (0.18, 0.4)$ for the Gegenbauer reconstruction. The Gegenbauer reconstruction appears exactly coincident with the exact result, unlike the direct reconstruction which shows distinctive Gibbs ringing. {\it Right panel:} relative error in the direct and Gegenbauer reconstructions, using two different truncation frequencies $\wmax = 25$ and $50$, again with $(\alpha, \beta) = (0.18, 0.4)$ for the Gegenbauer reconstruction. The Gegenbauer reconstruction incurs significantly smaller errors than the direct reconstruction.} 
\end{figure*}

Consider the function
\begin{align}
    f(t) =  \begin{cases}
        3e^{-t} & \text{ for } t > 0,\\~\\
        -2e^{2t} & \text{ for } t < 0,
    \end{cases}\label{eq:example_f}
\end{align}
which has a Fourier transform given analytically by 
\begin{align}
    \hat f(\omega) = \frac{4+5i\omega}{2\pi (2-i\omega + \omega^2)}. \label{eq:toy_example_modes}
\end{align}
$f(t)$ is analytic on each of $t < 0$ and $t > 0$, with a jump discontinuity at $t = 0$. In this example we will apply the Gegenbauer procedure to reconstruct $f(t)$ on the interval $t \in [0, 3]$ from the frequency modes in Eq.~\eqref{eq:toy_example_modes}.

For a given value of $\wmax$, the partial integrals $F(t; \wmax)$ are calculated by numerical integration for a discrete sample of times $t \in [0, 3]$. To project these partial integrals onto the Gegenbauer basis, it is necessary to first map the interval $[0,3]$ onto the interval $[-1,1]$ on which the Gegenbauer polynomials are defined. For a general interval $[a,b]$ this can be achieved using the affine transformation
\begin{align}
    t \in [a,b] \mapsto s(t) := \frac{2t - (a+b)}{b-a} \in [-1, 1].
\end{align}

The Gegenbauer coefficients $g_k^\lambda$ are computed up to some maximal value $k = N$ by numerical integration of Eq.~\eqref{eq:g_k_def} over $-1 \leq s \leq 1$, and the approximant
\begin{align}
    f(t) \approx \mathcal G_N\left(s(t)\right) = \sum_{k=0}^N g_k^\lambda C_k^\lambda\left(s(t)\right) \quad  \left(t \in [0, 3]\right) \label{eq:geg_approximant_example}
\end{align}
is then constructed. 

This process involves two free parameters: the degree $N$ at which the Gegenbauer series is truncated, and the weight parameter $\lambda$, or equivalently the proportionality constants $\alpha$ and $\beta$ from Eq.~\eqref{eq:param_prop}. The left panel of Fig.~\ref{fig:toy_example} displays the Gegenbauer reconstruction of $f(t)$ with $N = 9$, $\lambda = 20$, and $\wmax = 50$ (corresponding to $\alpha = 0.18$ and $\beta = 0.4$). The Gegenbauer approximant is visually indistinguishable from the exact result throughout the interval, in stark contrast to the Gibbs-afflicted direct reconstruction (also plotted with frequency truncation $\wmax = 50$ for comparison). The performance of the Gegenbauer reconstruction becomes particularly apparent in the right panel of Fig.~\ref{fig:toy_example}, which displays the relative error in the Gegenbauer and direct reconstructions. For the Gegenbauer reconstruction with $\wmax = 50$, the relative error is below $10^{-8}$ throughout a large part of the interval, rising to only around $6.4 \times 10^{-6}$ at the $t = 3$ end point, and only $1.4 \times 10^{-6}$ in the $t = 0^+$ limit. This compares to errors consistently greater than $1\%$ for the equivalent direct reconstruction, rising to approximately $80\%$ at $t = 0^+$. The right panel also clearly demonstrates the significantly faster convergence of the Gegenbauer reconstruction compared to direct reconstruction. Increasing $\wmax$ from $25$ to $50$ [with $\alpha = 0.18$ and $\beta = 0.4$ fixed and $N$ and $\lambda$ determined according to Eq.~\eqref{eq:param_prop}] results in only a small decrease in error when using the direct reconstruction method, compared to an approximately $4$ order of magnitude decrease in error throughout the interval when using the Gegenbauer procedure. 

\begin{figure}[htb]
  \centering
  \includegraphics[width=\linewidth]{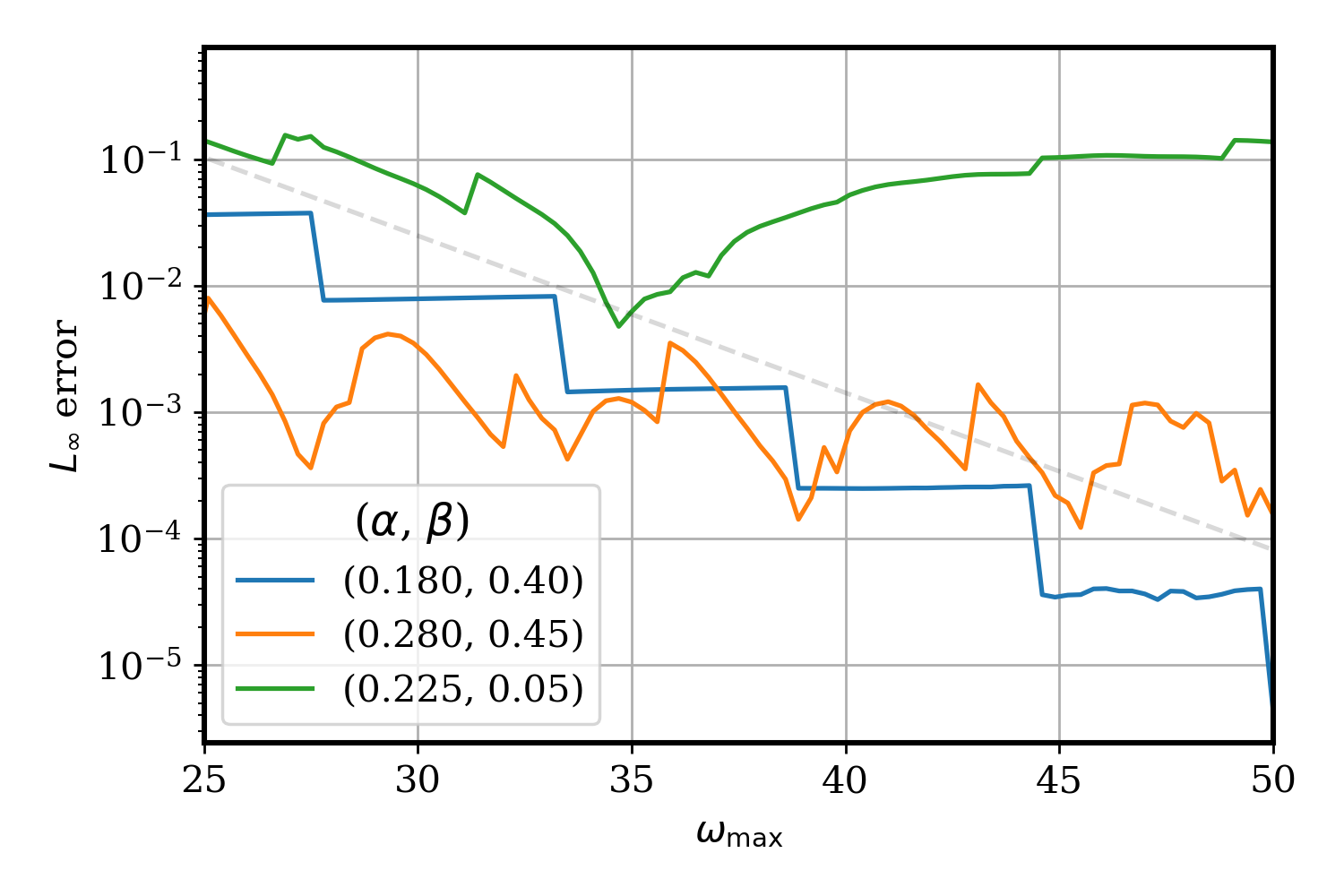}~\\~\\
  \caption{\label{fig:toy_example_err_decay} Decay of the $L_{\infty}$ error [as defined in Eq.~\eqref{eq:Linf_err}] in the Gegenbauer reconstructed value of $f(t)$ on $0 \leq t \leq 3$ as a function of $\omega_{\rm max}$ for three different choices of $(\alpha, \beta)$. For $(\alpha,\beta) = (0.18, 0.4)$ (blue) the error decays in a stepwise fashion, with jumps where the value of the discrete parameter $N$ increases. The envelope is clearly exponentially decaying, as illustrated by the exponential reference curve (grey dashed), confirming that the Gegenbauer reconstruction converges uniformly and exponentially. For $(\alpha, \beta) = (0.28, 0.45)$ (orange) the Gegenbauer reconstruction exhibits complicated, nonmonotone convergence, and the technique fails to converge entirely for $(\alpha, \beta) = (0.225, 0.05)$ (green).}
\end{figure}

Indeed, Fig.~\ref{fig:toy_example_err_decay} confirms the uniform exponential convergence of the Gegenbauer reconstruction method. In blue one can see the $L_{\infty}$ error in the Gegenbauer approximant~\eqref{eq:geg_approximant_example} on the interval $0 \leq t \leq 3$, i.e.,
\begin{align}
    \max_{0 \leq t \leq 3}\Big| 3e^{-t} - \mathcal{G}_N(s(t)) \Big|, \label{eq:Linf_err}
\end{align}
plotted as a function of $\wmax$ with the same fixed values of $(\alpha,\beta) = (0.18, 0.4)$ used in Fig.~\ref{fig:toy_example}. The $L_{\infty}$ error displays a distinctive stepwise decay, with an overall exponentially decaying envelope. The step-downs in the error occur at precisely the values of $\wmax$ at which $N$---the integer degree at which the Gegenbauer series in Eq.~\eqref{eq:geg_approximant_example} is truncated---jumps by $1$. Figure~\ref{fig:toy_example_err_decay} also demonstrates how the convergence rate depends on the values of $\alpha$ and $\beta$. For $(\alpha, \beta) = (0.28, 0.45)$, illustrated in orange, the error displays a complicated trend with features that cannot obviously be linked to the discrete jumps in $N$ this time. Despite this, an overall decay in the error from $\wmax = 25$ to $50$ is still clear, but it is not possible to definitively identify this decay as exponential in the circumstances. For $(\alpha, \beta) = (0.225, 0.05)$ (green) the situation is even more severe, as the error begins to increase with $\wmax$ after an initial decay, indicating a complete lack of convergence. 

It is clear then that the parameters $\alpha$ and $\beta$ must be carefully chosen to achieve optimal precision when using Gegenbauer reconstruction. Analytical schemes for selecting these parameters have been considered previously \cite{jackiewicz2004determination, FourGegParamOpt, gelb2005determining}, 
but we find them difficult to apply in practice. In Sec.~\ref{sec:scalar_field} we will empirically determine appropriate values for $\alpha$ and $\beta$ in our scalar-field reconstruction through numerical experimentation, and investigate the sensitivity of the reconstruction to our choice. In Sec.~\ref{sec:param_opt_discussion} we will then discuss possible parameter optimization strategies for actual self-force calculations where no comparison data are available. This will include a summary of various analytical approaches that have been suggested in the literature and their limitations. 

\section{Numerical demonstration for scalar-charge scattering}\label{sec:scalar_field}

In this section we apply the Gegenbauer reconstruction procedure to the calculation of the scalar-field modes $\psi_{\lm}(t,r)$ sourced by a scalar charge moving along the scattering geodesic illustrated in Fig.~\ref{fig:E1pt1_rmin4M_geodesic}. Specifically, we will calculate the mode $(\ell, m) = (10, 0)$ as a function of $t$ on the two surfaces $r_* = 5M$ and $r_* = 50M$, radii at which EHS reconstruction, respectively, can and cannot be relied upon to compute the field and its (internal) derivatives at the worldline. We verify the applicability of Gegenbauer reconstruction to both the internal region $r \leq r_p(t)$ and the external region $r \geq r_p(t)$, and confirm that the new approach converges exponentially as expected (for suitable choices of $\alpha$ and $\beta$). The Gegenbauer approach is validated against the EHS approach in situations where the latter is valid and reliable, and otherwise against results from the TD code. We will conclude the section by investigating the sensitivity of Gegenbauer reconstruction to the choice of parameters $N$ and $\lambda$, and assess the implications for future self-force calculations using this technique. The numerical methods used in this section to calculate frequency modes and perform the different methods of time-domain reconstruction are summarized in Appendix~\ref{app:numerical_methods}. A description of the TD code can be found in Sec. VIII of Ref.~\cite{BarackLong2022}. 

\subsection{Small-radius case}\label{sec:small_radius}
We begin by considering the reconstruction of the time-domain $(10,0)$ scalar-field mode on the surface $r_* = 5M$. Recalling Fig.~\ref{fig:direct_vs_ehs_vs_td}, there are three separate domains on this surface on which $\psi_{10,0}(t,r)$ and its derivatives are analytic: $t < -t_p$, $|t| < t_p$, and $t > t_p$, where $t = \pm t_p(r_*=5M) \approx \pm 13.0M$ are the two times at which the particle crosses the sphere $r_* = 5M$ (see Fig.~\ref{fig:int_ext_regions} for an illustration). For brevity, we focus our attention on reconstructing the fields in the vicinity of the outbound particle-crossing time, $t = t_p$, noting that reconstruction around $t = -t_p$ is nearly identical. We will also reconstruct only the mode $\psi_{10,0}$ and its time derivative $\partial_t\psi_{10,0}$ for simplicity. The Gegenbauer reconstruction technique can be applied equally well to the radial derivative $\partial_r\psi_{10,0}$ too.

We first seek to compute $\psi_{10,0}$ and $\partial_t\psi_{10,0}$ in the region $t > t_p$. This is an {\it internal} region, by which we mean that $5M < r_{*p}(t)$ for all $t > t_p$, so that the field points $(t, r_*=5M)$ lie ``inside'' the orbit (recall Fig.~\ref{fig:int_ext_regions}). More concretely, we will use Gegenbauer reconstruction to obtain these functions on a compact interval $J_I := [t_p, t_p+\Delta t]$. In the context of calculating the self-force using the mode-sum formula \eqref{eq:mode_sum_formula}, we are only interested in the value of $\psi_{10,0}$ and its derivatives at the end point $t = t_p^+$ itself, and so we are free to choose any value of $\Delta t>0$ which is convenient to us. Additionally, we must select the values of the weight parameter $\lambda$ and the degree $N$ at which the Gegenbauer polynomial expansion is truncated [this is equivalent to choosing the values of the proportionality constants $\alpha$ and $\beta$ defined in Eq.~\eqref{eq:param_prop}, given a fixed frequency truncation $\wmax$].

Unless otherwise stated, throughout Sec.~\ref{sec:scalar_field} we use a frequency truncation $\wmax = \wmax^{\rm VoP}$ for both direct and Gegenbauer reconstruction, where $\wmax^{\rm VoP}$ is the largest frequency at which we stored the inhomogeneous frequency modes $\psi_{\lmw}$ [calculated from the variation of parameters (VoPs) formula, Eq.~\eqref{eq:VoPRadialEqnScattering}]. Likewise, the EHS reconstruction is truncated at $\wmax = \wmax^{\rm EHS} = 1.945/M$, the largest frequency at which we store the EHS frequency modes \eqref{eq:int_ehs_def}. As outlined in Appendix~\ref{app:fd_calc_method}, we are unable to increase $\wmax^{\rm EHS}$ beyond this value due to numerical issues that affect the calculation of the EHS modes. The numerical calculation of the inhomogeneous modes is not affected by this issue, so we are free to calculate higher-frequency modes. For convenience, we opt to truncate at $\wmax^{\rm VoP} = 4.44625/M$, which is equal to $2001$ steps of size $M\Delta\omega = 1.25 \times 10^{-3}$ beyond $\wmax^{\rm EHS}$. There are no specific numerical barriers that would prevent us from increasing $\wmax^{\rm VoP}$ even further, besides the increased computational cost.

The parameters $(\Delta t, N, \lambda)$ for the Gegenbauer reconstruction of $\psi_{10,0}$ and $\partial_t\psi_{10,0}$ are chosen separately. In both cases we do this by performing the reconstruction using all combinations $(\Delta t, N, \lambda)$ from 
\begin{align}
    \Delta t &\in \{5M,\>10M,\>15M,\>20M,\>30M\},\label{eq:DT_range}\\
    N &\in \{1,\>2,\>...\>,\>9\}, \label{eq:N_range}\\
    \lambda &\in \{0.5,\>0.6,\>...\>,\>9.9,\>10.0\},\label{eq:lam_range}
\end{align}
and compute the $L_{\infty}$ error on $J_I$ in each case. The error is calculated by comparing with the results from EHS reconstruction, which is assumed to be exact. The ``optimal" parameters $(\Delta t^*, N^*, \lambda^*)$ are then chosen to be those which minimize the weighted average
\begin{align}
 \frac{\epsilon(\Delta t, N, \lambda-0.1) + 2\epsilon(\Delta t, N, \lambda) + \epsilon(\Delta t, N, \lambda+0.1)}{4}, \label{eq:avgd_error}
\end{align}
where $\epsilon(\Delta t, N, \lambda)$ is the $L_{\infty}$ error in the reconstruction with parameters $(\Delta t, N, \lambda)$. The weighted average is adopted to reduce the chance that we select an unrealistically fine-tuned set of parameters for which the error increases significantly under small perturbations to the continuous parameter $\lambda$. We emphasize that, while it is not possible to perform this optimization process in general applications, where comparison data are not presumed to be available, the purpose of this section is to demonstrate the capabilities of the Gegenbauer reconstruction method. The impact of suboptimal parameter choices on the accuracy of reconstruction will be assessed in Sec.~\ref{sec:parameter_choices}.

\begin{figure*}[tb!]
  \centering
  \includegraphics[width=0.9\linewidth]{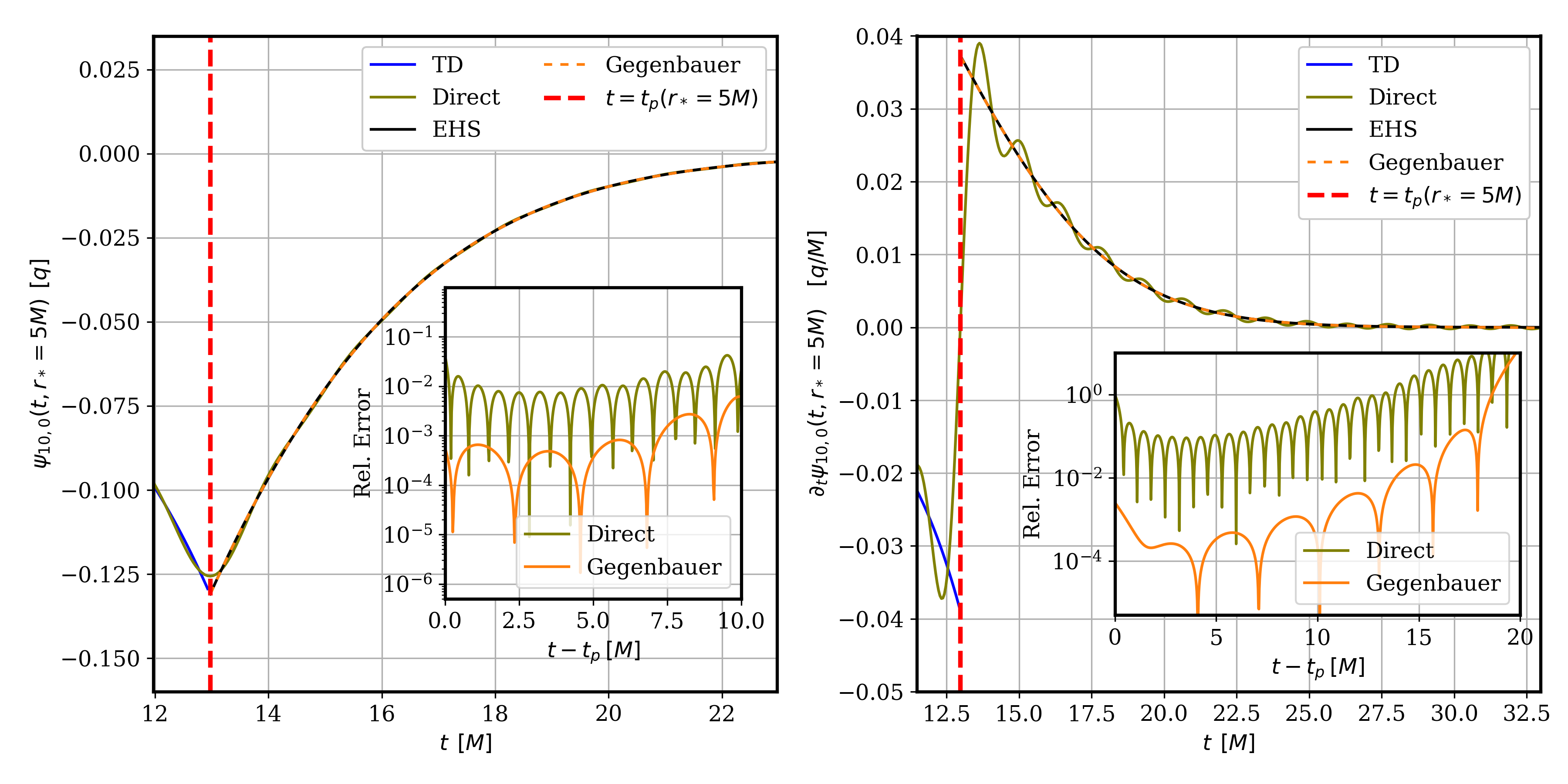}~\\~\\
\caption{\label{fig:l10m0_rstar5M_int_reconstructions}Comparison between the values of $\psi_{10, 0}(t, r_*=5M)$ ({\it left}) and $\partial_t\psi_{10,0}(t, r_*=5M)$ ({\it right}) in the internal region $t > t_p$, as obtained using direct (olive), EHS (black) and Gegenbauer (orange) reconstruction. The direct and Gegenbauer approaches make use of a frequency truncation $\wmax^{\rm VoP} = 4.44625/M $, while the EHS reconstruction is truncated at $\wmax^{\rm EHS} = 1.945/M$. The values of $\Delta t$, $N$ and $\lambda$ used for the Gegenbauer reconstruction are given for both the field mode and the time derivative in Table~\ref{tab:rstar_5M_opt_params}. The result of the TD code (blue) is included for $t < t_p$ to illustrate the behavior of the true fields across $t = t_p$ (indicated by the dashed red line). {\it Insets:} relative error compared to the EHS result for both the direct and Gegenbauer approaches. The Gegenbauer procedure outperforms direct reconstruction consistently throughout the reconstruction intervals, with a particularly large improvement for $\partial_t\psi_{10,0}$ in the vicinity of $t = t_p^+$.}
\end{figure*}

\begin{table}
\begin{ruledtabular}
\begin{tabular}{c c c c c}
 Quantity & Region & $\Delta t^*$ & $N^*$ & $\lambda^*$ \\
 \hline\vspace{-7pt}\\
 $\psi_{\lm}(t,r_*=5M)$ & Internal & $10M$ & $5$ & $4.1$\\
 $\partial_t\psi_{\lm}(t,r_*=5M)$ & Internal & $20M$ & $7$ & $3.8$\\
 $\psi_{\lm}(t,r_*=5M)$ & External & $20M$ & $10$ & $3.6$\\
 $\partial_t\psi_{\lm}(t,r_*=5M)$ & External & $20M$ & $9$ & $4.7$
\end{tabular} 
\end{ruledtabular} 
\caption{Optimal choices of the parameters $(\Delta t, N, \lambda)$ for the different fields and regions of interest on the spatial slice $r_* = 5M$, as calculated using the grid search described in Sec.~\ref{sec:small_radius}.
\label{tab:rstar_5M_opt_params}}
\end{table}

The optimal values of the parameters $(\Delta t, N, \lambda)$ for the reconstruction of $\psi_{10,0}$ and $\partial_t\psi_{10,0}$ in the vicinity of $t = t_p^+$ are given in the first two rows of Table~\ref{tab:rstar_5M_opt_params}.  The resulting reconstructions of the mode and its derivative are illustrated in Fig.~\ref{fig:l10m0_rstar5M_int_reconstructions}, where they are compared to the results of direct and EHS reconstruction. It is observed that the Gegenbauer and EHS results for $\psi_{10,0}$ and $\partial_t\psi_{10,0}$ are visually indistinguishable throughout their reconstruction intervals $J_I$. This contrasts sharply with the result obtained by direct reconstruction, which shows large-amplitude oscillations around the EHS/Gegenbauer results for the derivative $\partial_t\psi_{10,0}$. The direct result for $\psi_{10,0}$ itself is visually more faithful to the EHS result, but the direct curve has a distinctly smooth, rounded turning point in place of the true V-shaped kink in the field mode at $t = t_p$. Gegenbauer reconstruction, on the other hand, correctly recovers the straight $t > t_p$ side of the kink. The improved performance of Gegenbauer reconstruction compared to direct reconstruction is particularly apparent in the insets, which display the relative errors (compared to EHS reconstruction) for both approaches as functions of time. For $\psi_{10, 0}$, the Gegenbauer approach is around 1 order of magnitude more accurate than direct reconstruction throughout most of the interval, rising to around 2 orders of magnitude at $t = t_p^+$. For the derivative, the improvement compared to direct reconstruction is around 2 orders of magnitude throughout most of the interval. In the $t = t_p^+$ limit, where direct reconstruction fails to converge to the correct one-sided value for the derivative entirely, the Gegenbauer approach recovers the derivative with a relative error of only $2.5 \times 10^{-3}$.

\begin{figure}[H]
  \centering
  \includegraphics[width=\linewidth]{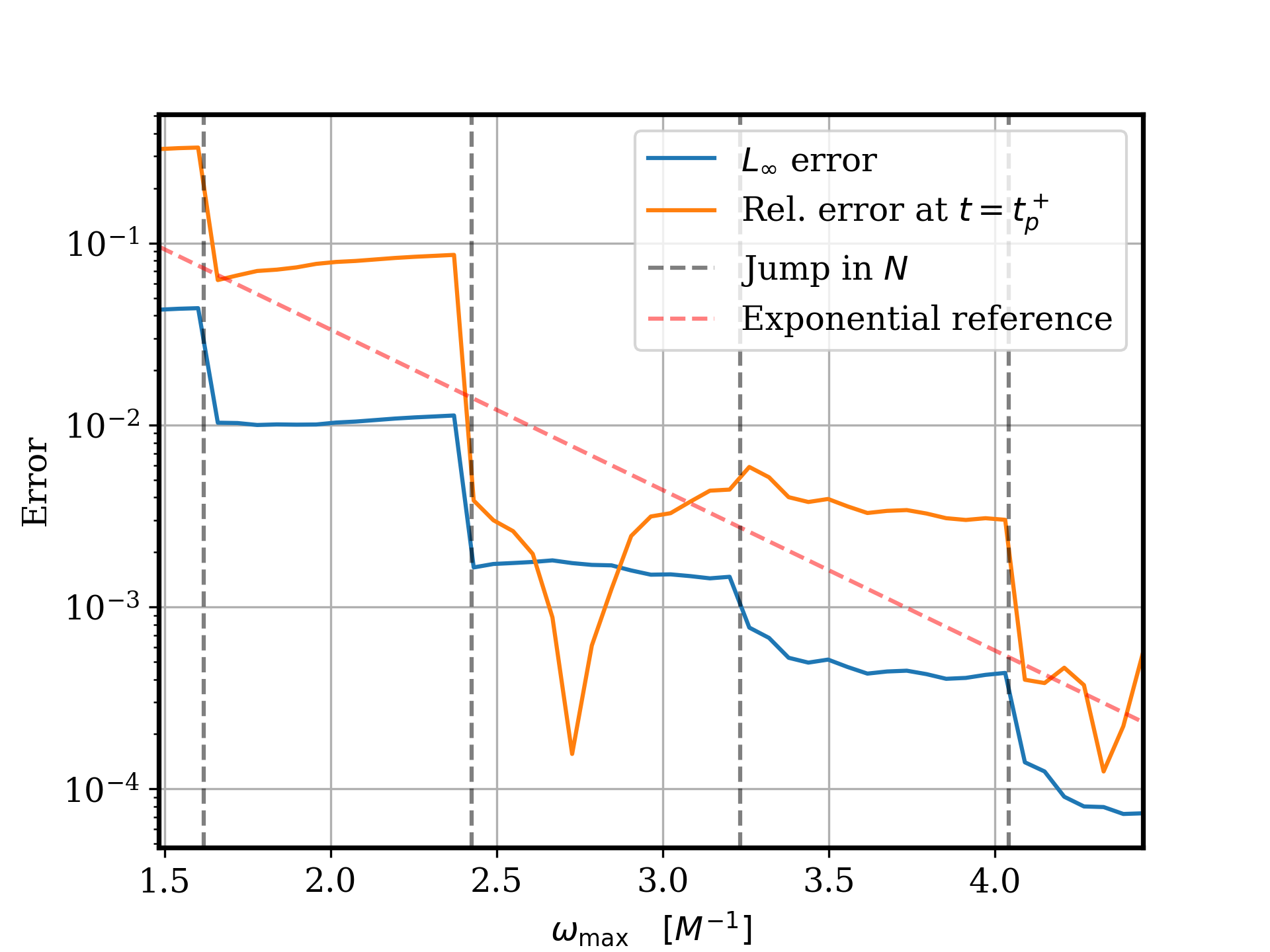}~\\~\\
  \caption{\label{fig:l10m0_convergence} 
  Decay of the $L_{\infty}$ error on $J_I$ (blue) and the relative error at $t = t_p^+$ (orange) as functions of $\wmax$ for the reconstruction of $\psi_{10, 0}(t, r_*=5M)$ on the interval $J_I = [t_p, t_p + 10M]$ using Gegenbauer reconstruction. The $L_{\infty}$ error decays stepwise in this example, jumping downwards when the value of $N$ increases (at the frequencies indicated by the vertical dashed grey lines), with an overall envelope which decays exponentially. An exponentially decaying reference curve which bounds the $L_{\infty}$ error is included for comparison (dashed red). The relative error, although it shows a more complicated behavior, decays at $t_p^+$ at approximately the same rate as the $L_{\infty}$ error when averaged across the range of $\wmax$. }
\end{figure}

We also take this opportunity to verify that the Gegenbauer reconstruction procedure converges exponentially when the values of $N$ and $\lambda$ grow linearly with the frequency truncation $\wmax$. Figure~\ref{fig:l10m0_convergence} displays both the $L_{\infty}$ error and the relative error at $t = t_p^+$ in the value of $\psi_{10,0}(t, r_*=5M)$ obtained by Gegenbauer reconstruction, as a function of the truncation frequency $\wmax$. The $L_{\infty}$ error displays the same exponential-bounded, stepwise decay that we observed for the analytical toy example in Fig.~\ref{fig:toy_example_err_decay}, confirming once again that Gegenbauer reconstruction converges uniformly and exponentially even in the case of a continuous frequency spectrum. The relative error at $t = t_p^+$ also decays in a nonmonotone fashion, which is, nonetheless, broadly consistent with the decay rate in the $L_{\infty}$ error across the sampled range of $\wmax$. Figure~\ref{fig:l10m0_convergence} was created by varying the truncation degree $N$ and weight parameter $\lambda$ with $\wmax$ according to Eq.~\eqref{eq:param_prop}, keeping the interval width fixed at the value $\Delta t^*$ given in Table~\ref{tab:rstar_5M_opt_params}. The constants
\begin{align}
    \alpha = \frac{2N^*+1}{2\wmax^{\rm VoP}}, \quad \beta = \frac{\lambda^*}{\wmax^{\rm VoP}}\label{eq:chosen_alpha_beta}
\end{align}
were chosen such that $N$ and $\lambda$ coincide with the optimal values $N^*$ and $\lambda^*$ from Table~\ref{tab:rstar_5M_opt_params} when $\wmax = \wmax^{\rm VoP}$. Note that the choice of $\alpha$ is nonunique due to the floor function $\lfloor.\rfloor$ appearing in Eq.~\eqref{eq:param_prop}. Indeed, any choice 
\begin{align}
    \frac{N^*}{\wmax^{\rm VoP}} \leq \alpha < \frac{N^*+1}{\wmax^{\rm VoP}} \label{eq:alpha_range}
\end{align}
will recover $N = N^*$ when $\wmax = \wmax^{\rm VoP}$. We arbitrarily choose the midpoint in Eq.~\eqref{eq:chosen_alpha_beta} to obtain a concrete value for $\alpha$ with which we could illustrate the convergence.
 
\begin{figure*}[!htb]
  \centering
  \includegraphics[width=0.9\linewidth]{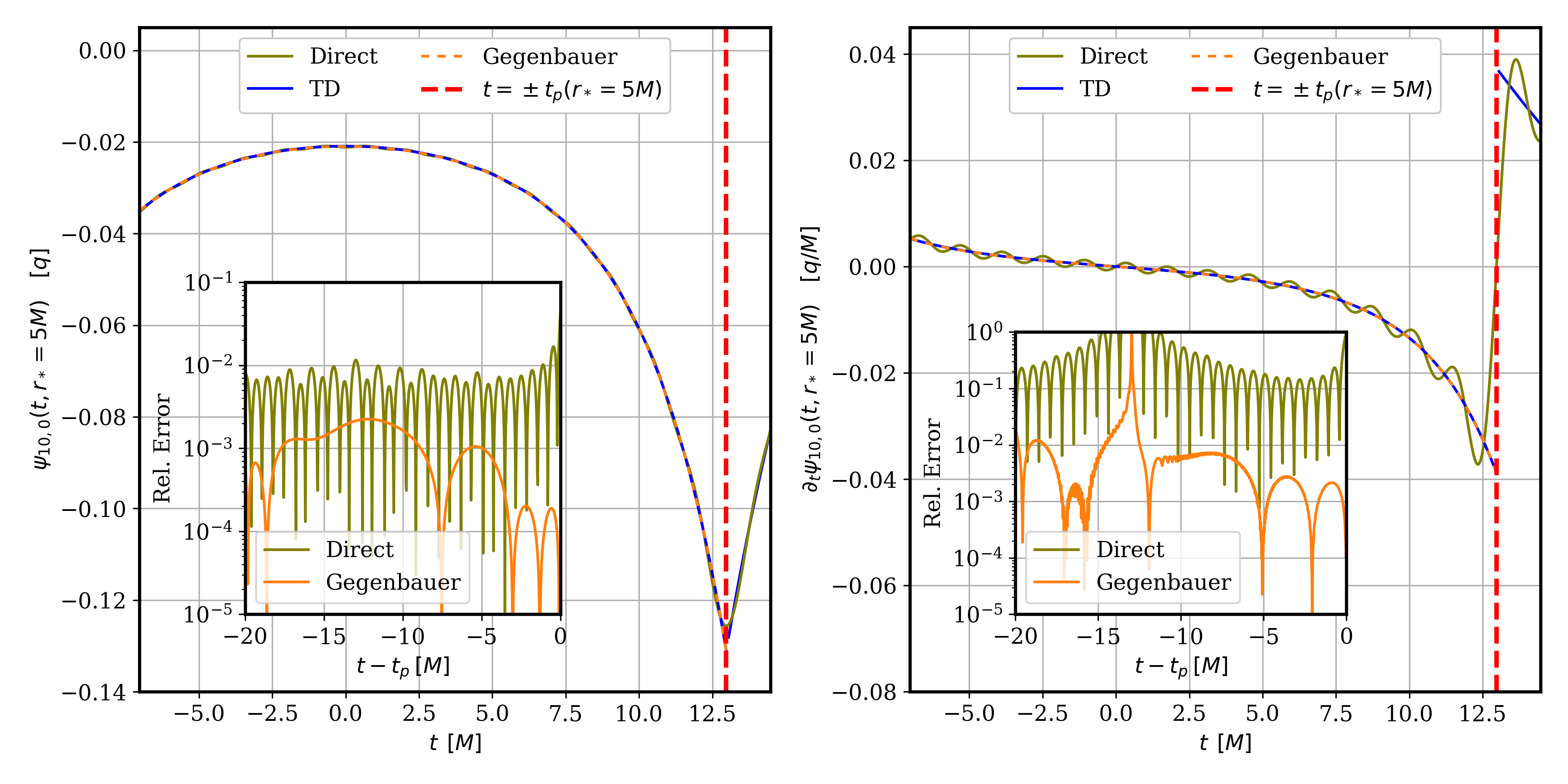}~\\~\\
  \caption{\label{fig:l10m0_rstar5M_ext_reconstructions}
  Same as Fig.~\ref{fig:l10m0_rstar5M_int_reconstructions}, but for the external region $|t| < t_p$. The TD code (blue) is used as a reference instead of the EHS, which is inapplicable in the external region.   The Gegenbauer procedure once again outperforms direct reconstruction, especially for $\partial_t\psi_{10,0}$. 
  }
\end{figure*}

Finally, we turn our attention to reconstructing $\psi_{10,0}(t, r_*=5M)$ and $\partial_t\psi_{10,0}(t, r_*=5M)$ in the vicinity of $t = t_p^-$. To be precise, we consider reconstruction on an interval $J_E := [t_p-\Delta t, t_p]$, where $\Delta t \leq 2t_p$ to ensure we exclude the inbound particle crossing time, $t = -t_p$.  Recalling Fig.~\ref{fig:int_ext_regions}, $r_p^*(t) < 5M$ for all $t \in J_E$, and hence this is an example of reconstructing the field at points in the {\it external} region, outside the particle's orbit. In this case, the physical time-domain modes and their derivatives cannot be reconstructed from the internal EHS given in Eq.~\eqref{eq:int_ehs_def}, so we use results from the TD code as our reference values instead. The optimal parameters $(\Delta t^*, N^*, \lambda^*)$ for the Gegenbauer reconstruction are obtained using the same grid search approach used for the internal reconstruction example, except we will calculate the error using the TD results instead of the EHS. We also modify the search space slightly, testing values
\begin{align}
    \Delta t &\in \{5M, 10M, 15M, 20M, 2t_p \},\\
    N &\in \{5,\>6,\>...\>,\>14\},
\end{align}
with $\lambda$ again selected from set \eqref{eq:lam_range}.
The upper limit on $\Delta t$ was reduced due to the condition $\Delta t \leq 2t_p$, and the upper limit on $N$ was increased after the algorithm initially selected optimal values $N^*$ at the upper end of the original range given in \eqref{eq:N_range}. 

The optimal values $(\Delta t^*, N^*, \lambda^*)$ for the reconstruction of the mode and its derivative in the external region near $t = t_p^-$ are listed in the final two rows of Table~\ref{tab:rstar_5M_opt_params}. The corresponding reconstructions of $\psi_{10, 0}$ and $\partial_t\psi_{10,0}$ are displayed in Fig.~\ref{fig:l10m0_rstar5M_ext_reconstructions}, alongside the results of direct reconstruction and the TD code. Once again the Gegenbauer reconstructions for both $\psi_{10,0}$ and $\partial_t \psi_{10,0}$ are visually coincident with the reference (TD) result throughout their respective reconstruction intervals, unaffected by the Gibbs ringing prominent in the direct reconstruction of $\partial_t \psi_{10,0}$. Examining the insets, which show the relative error in the Gegenbauer and direct reconstructions as compared to the result from the TD code, we see that Gegenbauer reconstruction is around $1$ order of magnitude more accurate than direct reconstruction throughout most of the interval. Gegenbauer reconstruction accurately obtains the value of $\psi_{10,0}$ at $t = t_p$ to within a relative error of $7.2\times 10^{-5}$, versus $4.2 \times 10^{-2}$ for direct reconstruction. Gegenbauer reconstruction is also consistently $1$ to $2$ orders of magnitude more accurate than direct reconstruction for the derivative $\partial_t\psi_{10,0}$ throughout most of the interval, correctly obtaining the $t = t_p^-$ limit of the derivative to within a relative error of $9.8 \times 10^{-4}$ compared to the TD code. Significantly, these tests confirm that Gegenbauer reconstruction presents a viable approach for reconstructing the time-domain modes and their derivatives accurately in the external region, where the EHS technique cannot be applied for scattering orbits.

\subsection{Large-radius case}
\label{sec:large_radius}

We will now consider the reconstruction of $\psi_{10,0}$ and $\partial_t\psi_{10,0}$ on the surface $r_* = 50M$. Due to the cancellation problem, the EHS approach cannot be used in practice to calculate the $\ell=10$ contribution to the self-force mode sum \eqref{eq:mode_sum_formula} at this radius along the scattering geodesic in Fig.~\ref{fig:E1pt1_rmin4M_geodesic}. 
Our goal is to demonstrate that Gegenbauer reconstruction remains effective at these larger radii where EHS fails, and presents a superior alternative to direct reconstruction in this region.

Considering first reconstruction in the internal region $t > t_p(r_* =50M) \approx 118M$, we select optimal parameters for the reconstruction of both $\psi_{10,0}$ and $\partial_t\psi_{10,0}$ using the grid search algorithm from Sec.~\ref{sec:small_radius}, with search space
\begin{align}
    \Delta t &\in \{5M, 10M, 15M, 20M, 30M, 50M, 75M, 100M \},\\
    N &\in \{1, 2, ..., 14\},\label{eq:N_range_IntDir50M}
\end{align}
and the same range \eqref{eq:lam_range} for $\lambda$. The new ranges for $\Delta t$ and $N$ were selected after initial tests favored values at the upper limit of the original ranges \eqref{eq:DT_range} and \eqref{eq:N_range}. In the absence of a reliable EHS result against which to compare, we calculate the $L_{\infty}$ error using the result from the TD code, which we now assume to be exact. 

\begin{table}
\begin{ruledtabular}
\begin{tabular}{c c c c c}
 Quantity & Region & $\Delta t^*$ & $N^*$ & $\lambda^*$ \\
 \hline\vspace{-7pt}\\
 $\psi_{\lm}(t,r_*=50M)$ & Internal & $50M$ & $11$ & $4.2$\\
 $\partial_t\psi_{\lm}(t,r_*=50M)$ & Internal & $30M$ & $7$ & $5.6$\\
 $\psi_{\lm}(t,r_*=50M)$ & External & $10M$ & $3$ & $12.4$\\
 $\partial_t\psi_{\lm}(t,r_*=50M)$ & External & $10M$ & $3$ & $3.7$
\end{tabular} 
\end{ruledtabular} 
\caption{Optimal choices of the parameters $(\Delta t, N, \lambda)$ for the different fields and regions of interest on the spatial slice $r_* = 50M$, as calculated using the grid search algorithm.}
\label{tab:rstar_50M_opt_params}
\end{table}

\begin{figure*}[tb!]
  \centering
  \includegraphics[width=0.9\linewidth]{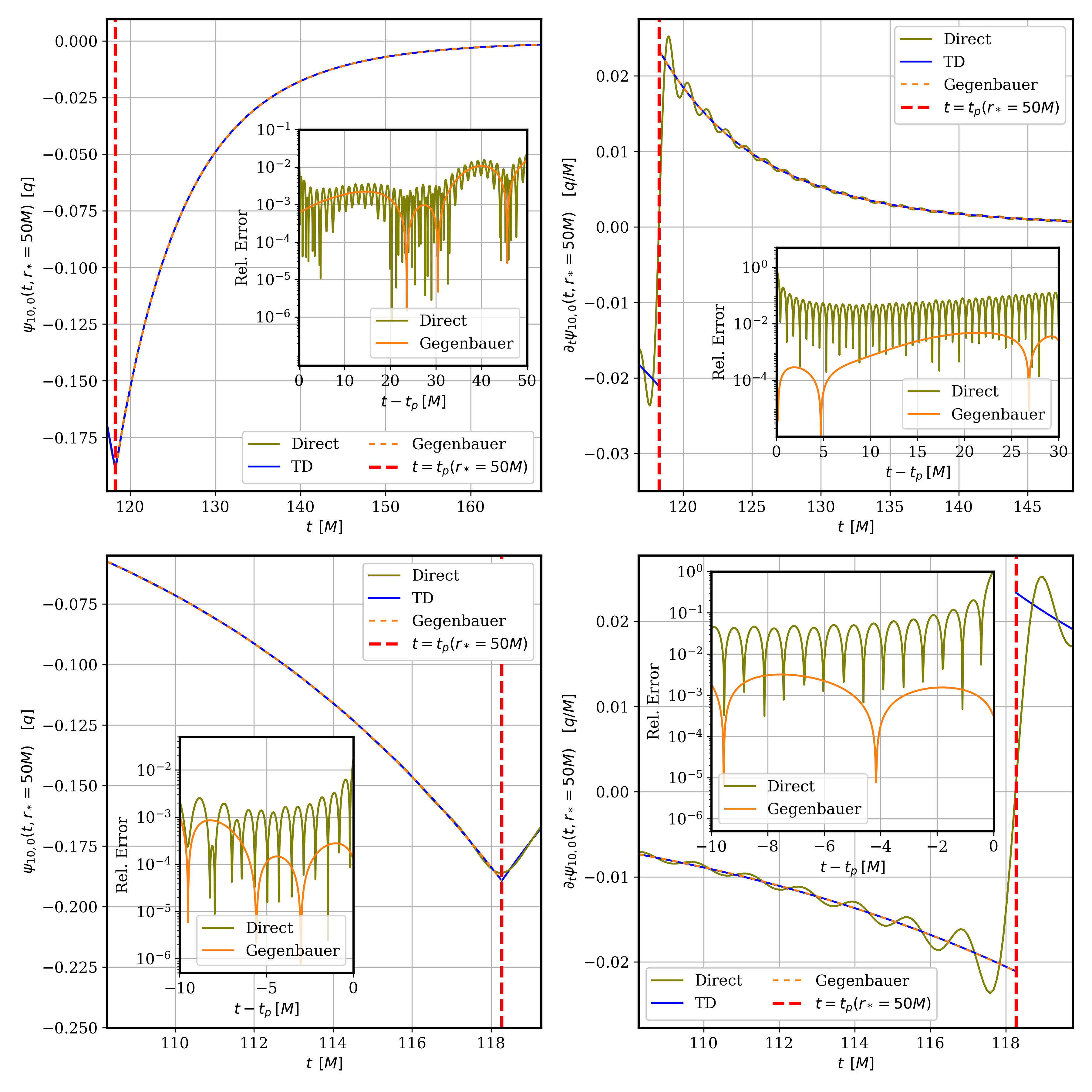}~\\~\\
  \caption{\label{fig:l10m0_50M_reconstructions} Similar to Figs.~\ref{fig:l10m0_rstar5M_int_reconstructions} and \ref{fig:l10m0_rstar5M_ext_reconstructions}, but this time showing the internal and external reconstructions for $\psi_{10,0}(t, r_* = 50M)$ and $\partial_t\psi_{10,0}(t, r_* = 50M)$. The TD code is used as reference, and the optimal parameters used for Gegenbauer reconstruction are given in Table~\ref{tab:rstar_50M_opt_params}. {\it Top left:} $\psi_{10, 0}$ in the internal region. {\it Top right:} $\partial_t\psi_{10, 0}$ in the internal region.  {\it Bottom left:} $\psi_{10, 0}$ in the external region.  {\it Bottom right:} $\partial_t\psi_{10, 0}$ in the external region. {\it Insets:} relative errors compared to the TD result. Gegenbauer reconstruction remains highly effective at $r_* = 50M$, without any evidence of degradation at larger radii.}
\end{figure*}

The resulting optimal values $(\Delta t^*, N^*, \lambda^*)$ are given in the first two rows of Table~\ref{tab:rstar_50M_opt_params}, and the corresponding reconstructions of $\psi_{10,0}$ and $\partial_t\psi_{10,0}$ are displayed in the top row of Fig.~\ref{fig:l10m0_50M_reconstructions}. Beginning with the mode $\psi_{10, 0}$ in the top left, the results of both Gegenbauer and direct reconstruction appear essentially indistinguishable by eye from the reference result computed using the TD code. Gegenbauer reconstruction outperforms direct reconstruction by a factor of between $2$ and $10$ at most times. When the relative error in the direct reconstruction starts to grow as $t \rightarrow t_p^+$ (reaching a value of $1.7 \times 10^{-2}$ at $t = t_p^+$), the Gegenbauer reconstruction remains faithful to the TD result, with a relative error of only $6.1 \times 10^{-4}$ at $t = t_p^+$. This is crucial for self-force calculations, where the accurate $t = t_p$ limits are required as input to the mode sum \eqref{eq:mode_sum_formula}. 
Looking next at the internal reconstruction of the derivative $\partial_t\psi_{10, 0}$, illustrated in the top right of Fig.~\ref{fig:l10m0_50M_reconstructions},  we see that Gegenbauer reconstruction once again remains visually coincident with the reference TD result, while direct reconstruction displays its characteristic Gibbs ringing and nonconvergence in the limit $t \rightarrow t_p^+$. Gegenbauer reconstruction outperforms direct reconstruction by $1$ to $2$ orders of magnitude away from $t = t_p$, and by an even greater margin in the $t \rightarrow t_p^+$ limit. The relative error in the Gegenbauer-reconstructed value of $\partial_t\psi_{10,0}$ at $t = t_p^+$ is $1.1 \times 10^{-4}$.

Considering now the external region $|t| < t_p$, we search for the optimal parameters for the reconstruction of $\psi_{10, 0}$ and $\partial_t\psi_{10,0}$, searching through values from \eqref{eq:DT_range}, \eqref{eq:N_range}, and
\begin{align}
    \lambda \in \{0.5, 0.6, ..., 14.9, 15.0\}.
\end{align}
This increased range for $\lambda$ was chosen after the optimal value for the reconstruction of $\psi_{10, 0}$ was initially found to lie at the upper end of the original range \eqref{eq:lam_range}. The optimal parameter choices $(\Delta t^*, N^*, \lambda^*)$ are given in the bottom two rows of Table~\ref{tab:rstar_50M_opt_params}, and the corresponding reconstructions are displayed in the second row of Fig.~\ref{fig:l10m0_50M_reconstructions}. The picture is much the same as in previous examples: Gegenbauer reconstruction agrees closely with the TD result by eye, for both $\psi_{10,0}$ and $\partial_t\psi_{10,0}$. The relative errors in $\psi_{10,0}$ are, except for a narrow region at the left-hand side of the reconstruction interval, mostly between a factor of 2 and a factor of 10 or more smaller for Gegenbauer reconstruction than for direct reconstruction. Gegenbauer reconstruction achieves a relative error in $\psi_{10,0}$ of only $1.3 \times 10^{-4}$ at $t = t_p^-$, compared to $1.7 \times 10^{-2}$ for direct reconstruction. The relative errors in $\partial_t\psi_{10,0}$ are mostly $1$ to $2$ orders of magnitude smaller for Gegenbauer reconstruction than for direct reconstruction. Gegenbauer reconstruction obtains the $t \rightarrow t_p^-$ limit of $\partial_t\psi_{10,0}$ correct to a relative error of $3.2 \times 10^{-4}$.

Altogether, we conclude that Gegenbauer reconstruction remains highly effective and superior to direct reconstruction at $r_*= 50M$, particularly in the $t \rightarrow t_p^\pm$ limits that are required for self-force calculations. This is especially true for the derivative $\partial_t\psi_{10,0}$, for which direct reconstruction does not converge to the correct one-sided limits. There is no evidence of any significant degradation in the accuracy of Gegenbauer reconstruction when increasing $r_*$ from $5M$ to $50M$, with the relative errors in $\psi_{10,0}$ and $\partial_t\psi_{10,0}$ in the $t \rightarrow t_p^\pm$ limits remaining at the $10^{-4} - 10^{-3}$ level in both cases. These results support the possible use of Gegenbauer reconstruction as an alternative to EHS reconstruction in the regime $r_p \gg \rmin$ where the latter fails.

\subsection{Sensitivity to choices of $\Delta t$, $N$ and  $\lambda$}\label{sec:parameter_choices}

In Secs.~\ref{sec:small_radius} and \ref{sec:large_radius} we demonstrated the precision of Gegenbauer reconstruction when the parameters $(\Delta t, N, \lambda)$ are chosen to be nearly optimal. The aim of this section is to explore the impact of nonoptimal parameter choices on the accuracy of reconstruction. 

\begin{figure*}[tbh]
  \centering
  \includegraphics[width=0.95\linewidth]{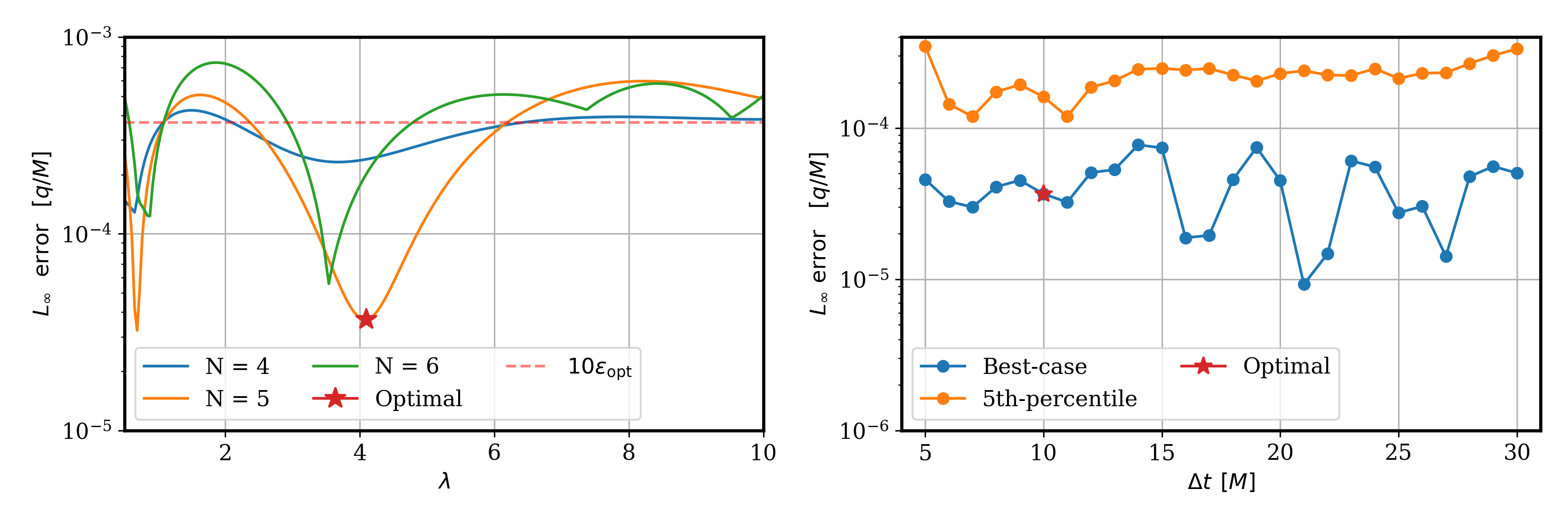}~\\~\\
  \caption{\label{fig:l10m0_rstar5M_IntField_sensitivity}Sensitivity of Gegenbauer reconstruction for the mode $\psi_{10,0}(t, r_*=5M)$ in the internal region $t > t_p$ to choices of the parameters $(\Delta t, N, \lambda)$. {\it Left panel:} $L_{\infty}$ error in the reconstructed value of $\psi_{10,0}$ on the reconstruction interval as a function of $\lambda$ for three discrete values $N \in \{N^*-1 = 4, N^* = 5, N^*+1 = 6\}$ and interval width $\Delta  t^* = 10M$. The red star marks the location of the ``optimal" parameters $(\Delta t^*, N^*, \lambda^*) = (10M, 5, 4.1)$ found in the previous section using a grid search, and the dashed red line indicates an error $10$ times that of the optimal parameter choice. {\it Right panel:} the best-case and $5$th-percentile $L_{\infty}$ errors (as defined in the text) for the reconstruction of $\psi_{10,0}(t, r_*=5M)$ on the interval $J_I := [t_p, t_p+\Delta t]$ as a function of interval width $\Delta t$. The red star once again marks the location of the ``optimal" parameters $(\Delta t^*, N^*, \lambda^*)$ used in the previous section.}
\end{figure*}

As a case study, we consider reconstructing $\psi_{10,0}$ on the surface $r_* = 5M$ in the internal region $t > t_p(r_*=5M)$ using different parameter combinations. In Sec.~\ref{sec:small_radius} we found the choice $(\Delta t^*, N^*, \lambda^*) = (10M, 5, 4.1)$ to be optimal for this example. The left panel of Fig.~\ref{fig:l10m0_rstar5M_IntField_sensitivity} displays the $L_{\infty}$ error in the Gegenbauer reconstruction of $\psi_{10,0}$ on the interval $[t_p, t_p+\Delta t^*]$ as a function of $\lambda$, for the optimal polynomial degree $N = N^* = 5$, and also for the surrounding values $N = 4$ and $6$. The error is computed by comparison with the result of the EHS code, which is assumed to be exact. We see that the $L_{\infty}$ error with the optimal parameters ($\epsilon_{\rm opt} := 3.7 \times 10^{-5}$) lies on a broad minimum of the $N = 5$ curve: the error is below $10^{-4}$ for all $3.35 \lesssim \lambda \lesssim 4.85$. Looking more widely, for $N = 5$ the error is below $10\epsilon_{\rm opt}$ (indicated by the dashed red line) for all $2.35 \lesssim \lambda \lesssim 6.2$. This means the error can be kept within a factor of $10$ of the optimal error with minimal tuning of $\lambda$ -- and to within a factor of a few using a slightly greater level of tuning -- provided the optimal value of $N$ is correctly identified. Choosing $N = 6$ instead, the error lies within a factor of $10$ of the optimal value for all $2.95 \lesssim \lambda \lesssim 4.75$, achieving a minimum error of $5.6\times 10^{-5}$ but staying below $10^{-4}$ only for a narrow range of $\lambda$. This means the error with $N = 6$ is typically larger than for $N = 5$, but can still be kept between a few and $10$ times the optimal error, albeit only by tuning $\lambda$ to a greater degree than is necessary for $N = 5$. On the other hand, for $N = 4$ the error is significantly flatter as a function of $\lambda$, so that the error lies at or below $10\epsilon_{\rm opt}$ for a wide range of $\lambda$, but generally stays at much higher levels than for $N = 5$ (the error drops only to $2.3\times 10^{-4}$ at the smooth minimum around $\lambda = 3.7$). The true minimum error for $N = 4$ occurs in the narrow feature at $\lambda \approx 0.65$, but we consider it unrealistic to attain this in real-world applications due to the fine-tuning required in $\lambda$. In this case, it appears that a suboptimal choice of $N$ (even with fine-tuning in $\lambda$) is likely to cause a larger reconstruction error than a minimally tuned choice of $\lambda$ with the optimal value of $N$.

We can also assess the impact of changing the width of the reconstruction interval, $\Delta t$. In the right panel of Fig.~\ref{fig:l10m0_rstar5M_IntField_sensitivity} we have plotted the ``best-case" (blue) and ``5th-percentile'' (orange) errors in the Gegenbauer reconstruction as a function of $\Delta t$. To obtain these errors for a given $\Delta t$, we perform Gegenbauer reconstruction with that value of $\Delta t$ and all combinations of $N$ and $\lambda$ from sets \eqref{eq:N_range} and \eqref{eq:lam_range} (the same ranges that were used for the grid search optimization for this example in Sec.~\ref{sec:small_radius}), and compute the $L_{\infty}$ errors for each. The best-case error for width $\Delta t$ is simply the $L_{\infty}$ reconstruction error with the parameter combination $(N^*(\Delta t), \lambda^*(\Delta t))$ which minimizes the weighted average \eqref{eq:avgd_error}, and the 5th-percentile error is the least value which is greater than the smallest $5 \%$ of (nonaveraged) $L_{\infty}$ errors from across the different $(N, \lambda)$ combinations sampled. Note that Fig.~\ref{fig:l10m0_rstar5M_IntField_sensitivity} includes $\Delta t$ values that were not included in set \eqref{eq:DT_range}, and the error corresponding to the ``optimal" parameters (indicated by the red star) found by grid search in Sec.~\ref{sec:small_radius} is not the lowest best-case error for the expanded set. In fact, the best-case error oscillates between approximately $10^{-5}$ and $8 \times 10^{-5}$, with the oscillations having greater amplitude in $\Delta t \gtrsim 10M$. This high variability in best-case error---a range of almost $1$ order of magnitude---underlines the potential benefit of treating $\Delta t$ as an optimizable parameter. On the other hand, the 5th-percentile errors, which give an idea of the error incurred when using less-than-optimal values of $(N, \lambda)$, display significantly less variation with $\Delta t$. 

In our analysis we have used $L_{\infty}$ as a conservative error norm. For self-force calculations, we require an accurate determination of $\psi_{\lm}(t,r)$ and its derivatives at the location of the particle and are less interested in what the reconstruction yields away from the particle. The local error at the particle is thus a more relevant measure. We have examined the local error in all the examples considered above, and found that it depends on the parameter choice in much the same way as the $L_{\infty}$ error. The behavior demonstrated in Fig.~\ref{fig:l10m0_rstar5M_IntField_sensitivity} represents well the behavior of the error at the particle as well. 

The above examples were found to be broadly representative of other scenarios we have analyzed in the preparation of this work. However, it is of course possible that other relevant examples of self-force calculations based on Gegenbauer reconstruction may show a different degree of sensitivity to the choice of parameters. More importantly, we have not yet tackled the fundamental question of how to best select these parameters in the absence of comparison data. Potential approaches to this problem will be discussed in the next section.

\section{Discussion}\label{sec:discussion}
In this work we demonstrated how the Gegenbauer reconstruction technique can be used to obtain with exponential, uniform accuracy the spherical harmonic modes $\psi_{\lm}(t,r)$ of the scalar field and its derivatives sourced by a scalar point charge scattered off a Schwarzschild black hole. Accurate knowledge of these modes is required to calculate the scalar-field self-force acting on such a particle using mode-sum regularization, the primary objective of the ongoing self-force scattering program. The significance of Gegenbauer reconstruction lies in the resolution of two key deficiencies of the standard EHS approach to scattering calculations. As demonstrated in Secs. \ref{sec:small_radius} and \ref{sec:large_radius}, and unlike the EHS approach, Gegenbauer reconstruction can be applied in both the internal $r < r_p(t)$ and external $r > r_p(t)$ regions of the spacetime for scattering orbits. Gegenbauer reconstruction also remains accurate at radii far from periapsis, avoiding the problem of severe numerical cancellation that afflicts the EHS at early and late times.

In Sec.~\ref{sec:geg_toy_example} we noted that it is important to carefully choose the Gegenbauer reconstruction parameters $(N, \lambda)$ [or equivalently the proportionality constants $\alpha$ and $\beta$ appearing in Eq.~\eqref{eq:param_prop}] in order to maximize the accuracy of our reconstruction. For the scalar-field examples in Sec.~\ref{sec:scalar_field}, we additionally treat the width $\Delta t$ as a customizable reconstruction parameter, and find approximately optimal values for all three parameters using a grid search algorithm to minimize an error metric based on the $L_{\infty}$ reconstruction error as measured against reference results generated using the EHS and TD codes. The sensitivity of the reconstruction error to variation of the reconstruction parameters is explored in Sec.~\ref{sec:parameter_choices}. Reconstruction with suboptimal parameters increases the reconstruction error, but our tests suggest this increase is not catastrophic, and errors within a factor of $10$ of the optimal value can typically be achieved across some range of parameter space sufficiently close to the optimal parameters. This increases our confidence that near-optimal reconstruction errors can still be achieved in real-world applications, where the parameters must, in general, be chosen without external reference data. Developing reliable, near-optimal parameter selection techniques is the most important challenge as we move toward the use of Gegenbauer reconstruction for full self-force calculations.

The remainder of this discussion is dedicated to the question of how to best extend our proof-of-concept spherical harmonic mode calculation into practical calculations of the self-force itself. We will first address potential solutions to the problem of parameter selection, before discussing broader considerations we must take when designing and performing our self-force calculation. Recognizing that our work represents only an initial and highly targeted exploration of the huge body of work seeking solutions to the Gibbs phenomenon, we will also consider some extensions and alternatives to the Gegenbauer reconstruction method that may form the basis of future work. 

\subsection{Approaches to on-the-fly parameter selection}
\label{sec:param_opt_discussion}

A very basic approach to parameter selection is to mimic the path taken in this work and use existing numerical results (obtained, for example, with an alternative time-domain code) to find combinations $(\Delta t^*, N^*, \lambda^*)$ which minimize an appropriate error metric. While this is clearly only possible for selected cases where comparison data already exist, it may be possible to use these selected results to derive empirical rules or interpolating formulae to predict the optimal parameters for arbitrary scattering systems. Leaving aside questions of whether this can actually be achieved in practice, the dependence on external data is fundamentally unsatisfactory. Relying on the existence of comparison data relegates Gegenbauer reconstruction to being a second-line method, which can only be used when alternative schemes are already available. This prevents us from independently tackling cutting-edge problems such as the calculation of the gravitational self-force along scattering orbits, for which no other approaches currently exist. Furthermore, even when comparison data exist, we may recover biased estimates for the optimal parameters if the error in the comparison data itself is comparable to, or greater than, the true optimal error in Gegenbauer reconstruction. This may in turn limit the accuracy of Gegenbauer reconstruction to around the same as that of the reference method. 

It would therefore be preferable to develop a parameter choice strategy that does not rely on external data. One way to do this might be to construct a hybrid EHS/Gegenbauer approach, which makes use of the EHS near to periapsis, and then switches to Gegenbauer reconstruction further away. The optimal Gegenbauer parameters could be determined in an overlap region where the EHS is still highly accurate, and then iteratively updated as we move out along the orbit. These updates could be based on internal convergence tests, varying the frequency truncation $\wmax$ to estimate the accuracy of different approximants and/or their rates of convergence, but this might prove challenging given the stepwise (or worse) convergence exhibited in Figs.~\ref{fig:toy_example_err_decay} and \ref{fig:l10m0_convergence}. It would also not be possible to use the EHS to calibrate Gegenbauer reconstruction in the external region.

Ideally, we would dispense with numerical calibration altogether and predict the optimal parameters {\it a priori} using analytical results instead. The question of how to optimize the Gegenbauer reconstruction parameters has been tackled for reconstruction from Chebyshev \cite{jackiewicz2004determination} and discrete Fourier \cite{FourGegParamOpt, gelb2005determining} coefficients. The common idea behind these methods is to obtain analytical bounds on the regularization and truncation errors in Eq.~\eqref{eq:err_bound}, and then minimize the truncation error subject to the condition that both errors are comparable in size. To be more precise, these approaches derive bounds \cite{jackiewicz2004determination, FourGegParamOpt, gelb2005determining}
\begin{align}
    \| f - \mathcal{F}_N\|_{\infty} &\leq A(m)\>q_R(\alpha,\beta)^m \label{eq:R_bound}\\
    \| \mathcal{F}_N - \mathcal{G}_N\|_{\infty} &\leq A(m)\>q_T(\alpha, \beta)^m, \label{eq:T_bound}
\end{align}
where $m$ is either the maximal Fourier mode number or the degree of the Chebyshev expansion (depending on the case under consideration), $A(m)$ is a constant or polynomial in $m$, and $\alpha$ and $\beta$ are the proportionality constants,
\begin{align}
    N = \lfloor\alpha m\rfloor, \qquad\lambda = \beta m.
\end{align}
The functions $q_R(\alpha,\beta)$ and $q_T(\alpha, \beta)$ are given analytically \cite{gottlieb1992gibbs, gottlieb1995gibbs, gottlieb1997gibbs} in terms of $\alpha$, $\beta$, and also a constant $\rho \geq 1$ (assumed to exist), satisfying
\begin{align}
    \max_{-1 \leq t \leq 1}\left|\frac{d^k f(t)}{dt^k} \right| \leq C_{\rho}\frac{k!}{\rho^k} \quad \forall\> k \geq 0,
\end{align}
where $C_{\rho}$ is independent of $k$ and the reconstruction interval is assumed to be $-1 \leq t \leq 1$ without loss of generality. The optimal parameters $(\alpha,\beta)$ are identified by minimizing $q_T(\alpha,\beta)$ subject to $q_T(\alpha,\beta) = q_R(\alpha, \beta)$.  We anticipate that bounds similar to \eqref{eq:R_bound} and \eqref{eq:T_bound} can be derived for the continuous Fourier reconstruction problem, allowing the application of a scheme along the lines of Refs.~\cite{jackiewicz2004determination, FourGegParamOpt, gelb2005determining}. To make this practical, we must be able to estimate the value of $\rho$ (or an equivalent quantity characterizing the analyticity of $f$ on the reconstruction interval \cite{jackiewicz2004determination, gelb2005determining}) from the Fourier transform $\hat f(\omega)$ (truncated at finite frequency). Two possible solutions to this problem in the discrete case are proposed in Ref.~\cite{gelb2005determining}, and we intend to explore their application to continuous spectra. 

\subsection{Extension to full self-force calculations}

There are several problems to tackle when extending the calculation of the field modes performed in this work to a calculation of the self-force itself. Gegenbauer reprojection carries a numerical overhead from having to calculate the partial Fourier integrals at discrete locations within a compact time interval for each radial position along the orbit, and then having to evaluate the integrals \eqref{eq:g_k_def} to project onto the Gegenbauer polynomial basis. This contrasts with the EHS approach, for which only a single Fourier integral is required at each position along the orbit. We expect that the impact of this overhead can be reduced using efficient quadrature rules. The current Gegenbauer reconstruction code is written in \texttt{Python} and would likely benefit from being rewritten in \texttt{C}, as is the case for the high-performance EHS code from Ref.~\cite{Whittall:2023xjp}. The cost in wall time can also be reduced by performing time-domain reconstruction at the different orbital positions and different $(\ell, m)$ modes in parallel. Finally, we note that EHS reconstruction also becomes inefficient at early and late times, when the high degree of cancellation in the EHS Fourier integrals \eqref{eq:intEHS_TD} causes the integrator to explore progressively smaller subintervals in a near-futile attempt to resolve the highly oscillatory integrand. This creates a massive increase in the cost of EHS reconstruction and thus decreases the marginal cost of using Gegenbauer reconstruction instead. 

In any case, the bottleneck in the current EHS approach for scattering lies in the pregeneration of the frequency-domain data described in Appendix~\ref{app:fd_calc_method}. Gegenbauer reconstruction is likely to exacerbate this issue for two reasons. First, we made use of larger  values of $\wmax$ for Gegenbauer reconstruction than for EHS reconstruction. For a constant frequency spacing, the cost of generating the frequency modes will increase faster than linearly in $\wmax$ due to the greater cost of computing higher-frequency modes, for which we must evaluate increasingly more oscillatory integrals. The latter effect can be minimized by using quadrature routines specially adapted to oscillatory integrands, as described in Sec. V B of Ref.~\cite{Whittall:2023xjp}. The frequency mode calculation is also highly parallelizable because the calculation of modes with different values of $\ell$ and $\omega$ are completely independent, allowing significant acceleration when running on a high-performance computing cluster. 

Second, Gegenbauer reconstruction uses the inhomogeneous frequency modes given by the variation of parameters formula in Eq.~\eqref{eq:VoPRadialEqnScattering}, which requires the evaluation of two integrals \text{per radial position} along the orbit. This contrasts with the EHS modes \eqref{eq:int_ehs_def}, for which only a single integral $C_{\lmw}^-$ is required for each orbit and choice of $(\ell, m, \omega)$ (for bound orbits a second normalization integral is also required to perform external reconstruction).  Fortunately, the integrals in Eq.~\eqref{eq:VoPRadialEqnScattering} depend on the evaluation radius $r$ only through the integration limits, so it should be possible to significantly reduce computational cost by reusing portions of the integrals or by caching values of the integrand. Alternatively, the integrals may be recast as ordinary differential equations and solved numerically to efficiently obtain the integrals as functions of their limits. 

\subsection{Alternative reconstruction methods}
In this work we utilized the Gegenbauer reprojection method as it was first introduced in Refs.~\cite{gottlieb1992gibbs, gottlieb1994resolution, gottlieb1995gibbs, gottlieb1997gibbs} in the 1990s. We will now briefly review some developments in the decades since, which aim to achieve even more accurate and robust reconstructions of discontinuous functions from their Fourier coefficients. Some of these may prove useful for self-force calculations in the future.

A known limitation of Gegenbauer polynomials as a Gibbs complementary basis originates in the requirement that $\lambda \rightarrow \infty$ in proportion to $\wmax$ in order to achieve exponential convergence. For large $\lambda$, the weight function $(1-s^2)^{\lambda-1/2}$ appearing in the inner product \eqref{eq:g_k_def} becomes narrowly concentrated around $s = 0$, so that the Gegenbauer approximant \eqref{eq:G_N_def} is inherently extrapolatory: the values of the approximant at the end points of the interval are determined from knowledge of the function in the central region only \cite{GELB20063}. This in turn leads to the ``generalized Runge phenomenon", in which the Gegenbauer approximant can converge poorly (or even diverge) at the end points of the interval unless the ratio $\lambda/N$ is kept sufficiently small \cite{BOYD2005253}. 

To resolve this issue, Ref.~\cite{GELB20063} introduced the {\it Freud} polynomials $F_n^\lambda(s)$ as an alternative Gibbs complementary basis. A generalization of Hermite polynomials, $F_n^\lambda(s)$ are orthogonal polynomials on $\mathbb{R}$ with respect to the weight function $\exp(-c s^{2\lambda})$, where $\lambda$ is now a positive integer, and $c$ is some chosen positive constant. Rather than concentrating at $s=0$ for large $\lambda$, the Freud weight function instead has the limiting form
\begin{align}
    \lim_{\lambda\rightarrow \infty} \exp\left(-cs^{2\lambda}\right) = 
    \begin{cases}
        1 & (|s| < 1),\\~\\
        0 & (|s| > 1),
    \end{cases}
\end{align}
which is the weight function for the Legendre polynomials. The additional property that the weight function converges to a weight whose orthogonal polynomials also obey condition (1) from the definition of a Gibbs complement in Sec.~\ref{sec:Geg_intro} leads to what Ref.~\cite{GELB20063} terms a ``robust Gibbs complement'', and resolves the generalized Runge phenomenon. Regrettably, it is challenging to evaluate Freud polynomials in general, and implementations do not exist in common numerical libraries such as the GNU Scientific Library \cite{GSL_lib}, \texttt{scipy} \cite{2020SciPy-NMeth} or {\it Mathematica}. 

There also exist different variants of the reprojection procedure. Rather than projecting the partial Fourier representation of the target function directly onto the Gegenbauer polynomials, Ref.~\cite{IPRM} introduced an approach where the projection is performed at the level of the basis elements themselves. More precisely, for a discrete Fourier expansion one formulates a linear system of equations relating the Fourier coefficients of the function to its polynomial expansion coefficients and a matrix of inner products between the elements of the two bases. Given finitely many Fourier coefficients, this linear system can be truncated and solved to obtain approximations to the leading coefficients of the (exponentially convergent) polynomial expansion. This technique, known as the inverse polynomial reconstruction method (IPRM) is not equivalent to the Gegenbauer procedure used in this paper \cite{IPRM}. It can be shown that the result of the IPRM is independent (up to numerical error) of the weight parameter $\lambda$ of the Gegenbauer polynomials used \cite{IPRM}, and the degree of the polynomial is fixed by the number of available Fourier coefficients. Parameter selection is therefore not an issue for the IPRM.

Unfortunately, however, the IPRM suffers from its own cancellation problem: the condition number of the linear system grows rapidly with the degree of the polynomial expansion \cite{IPRM, IPRM_general}, amplifying errors in the Fourier coefficients and causing issues with finite precision floating point arithmetic. In fact, it has been shown \cite{StabilityBarrier} in general that any exponentially convergent method for recovering an analytic, nonperiodic function from its discrete Fourier coefficients will inevitably be exponentially ill-conditioned as well.\footnote{This result also implies that the Gegenbauer reconstruction method is itself numerically unstable in the $\wmax \rightarrow \infty$ limit. This asymptotic ill-conditioning was not observed empirically at the finite values of $\wmax$ used during our numerical tests, and our numerical results in Sec.~\ref{sec:scalar_field} agreed closely with the reference data. The IPRM, on the other hand, suffers acutely with numerical cancellation in our practical experience.} 

\begin{figure}[t!]
  \centering
  \includegraphics[width=0.95\linewidth]{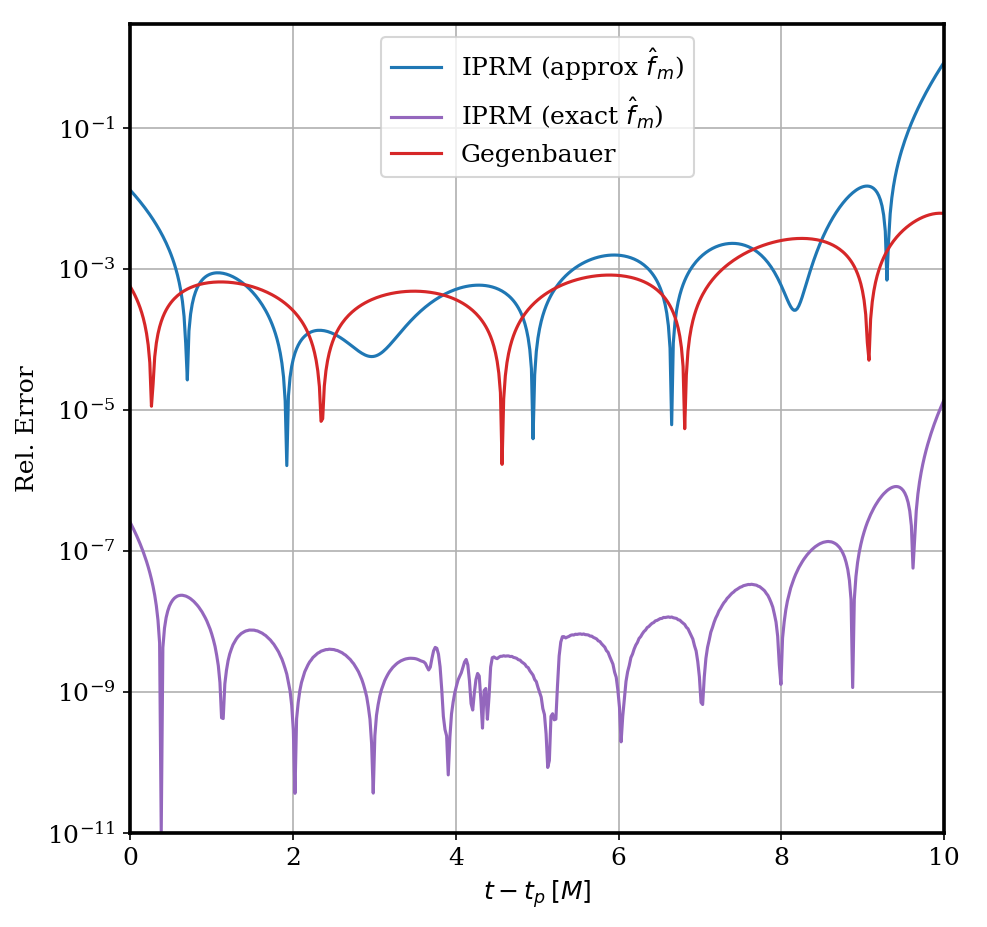}~\\~\\
  \caption{\label{fig:IPRM_demo}Relative error (compared to EHS reconstruction) in three different reconstructions of $\psi_{10, 0}(t, r_*=5M)$ on the interval $t \in [t_p, t_p+10M]$. Our Gegenbauer reconstruction (red) with the reconstruction parameters given in Table~\ref{tab:rstar_5M_opt_params} is compared with the IPRM using both ``approximate" (blue) and ``exact" (purple) discrete Fourier coefficients (as defined in the text). Both IPRM results include discrete Fourier modes up to and including $|m| = 5$ in Eq.~\eqref{eq:discrete_exp}.}
\end{figure}

More fundamentally, these inverse reconstruction methods have only been formulated for a discrete sampling basis, and do not extend to the continuous Fourier basis we use. One approach we considered to circumvent this issue is to first re-expand our fields in a discrete Fourier basis on some compact time interval. A function $f(t)$ with continuous Fourier transform $\hat f(w)$ can be expanded on $a \leq t \leq b$ in the form 
\begin{align}
    f(t) = \sum_{m = - \infty}^{+\infty} \hat f_m e^{-im\pi (t-Q)/P} ,\label{eq:discrete_exp}
\end{align}
where $P := (b-a)/2$, $Q := (b+a)/2$, and the discrete Fourier coefficients $\hat f_m$ are related to the continuous Fourier transform by
\begin{align}
    \hat f_m =\displaystyle\int_{-\infty}^{+\infty} d\omega\> \hat f(w) \>e^{-i\omega Q}\>\text{sinc}\left(m - \omega P/\pi\right), \label{eq:cts_to_discrete} 
\end{align}
with $\text{sinc}(x) := \sin(\pi x)/(\pi x)$. We used this relation to re-expand $\psi_{10, 0}(t,r_*=5M)$ in discrete Fourier harmonics on the time interval $t \in [t_p, t_p + 10M]$, and applied the IPRM to reconstruct the function from these discrete modes. Figure~\ref{fig:IPRM_demo} compares the precision of Gegenbauer reconstruction (with the optimal reconstruction parameters from Table~\ref{tab:rstar_5M_opt_params}) to two different applications of the IPRM: one (``approx $\hat f_m$") which makes use of Eq.~\eqref{eq:cts_to_discrete} with frequency truncation $\wmax = 4.44625/M$ to compute the discrete Fourier coefficients, and another (``exact $\hat f_m$") that obtains these coefficients by decomposing the EHS result itself into the discrete Fourier basis. We were forced to use a truncation of only $|m|_{\rm max} = 5$ in Eq.~\eqref{eq:discrete_exp} in order to keep the condition number of the linear system small. Despite this, using the approximate coefficients, the IPRM is comparable in accuracy to Gegenbauer reconstruction at most times except for the end points, where Gegenbauer reconstruction performs significantly better. On the other hand, the IPRM significantly outperforms Gegenbauer reconstruction everywhere when the exact discrete Fourier coefficients are used, achieving a relative precision of around $2 \times 10^{-7}$ at $t = t_p^+$, much lower than any other error observed in this work. Clearly, our approximate mapping between the continuous and discrete Fourier expansions limits the potential of the inverse reconstruction methods.

Overall, Gegenbauer reconstruction remains the most accurate and reliable reprojection approach for continuous spectra that we have considered so far. Nevertheless, it is prudent to remain open to alternatives, especially those that do not require careful parameter selection or that might achieve significantly greater numerical precision or efficiency. We acknowledge that there are many approaches to accurate Fourier reconstruction that we have not yet explored, including some techniques specifically tailored toward the continuous-spectrum case \cite{ContinuousGS}. We also highlight that the techniques discussed in this work may be useful for the highly eccentric, bound-orbit self-force problem, which also suffers from severe numerical cancellation when using the EHS approach \cite{vandeMeent2016}. In particular, the bound-orbit problem has a naturally discrete spectrum, allowing the application of the IPRM methods without the need for any problematic re-expansion. 

\subsection{Outlook}
Our next step is to construct an EHS/Gegenbauer hybrid along the lines described in Sec.~\ref{sec:param_opt_discussion}, with the goal of extending the scalar-field self-force calculation of Ref.~\cite{Whittall:2023xjp} to larger radii, enabling more accurate calculations of the scatter angle. In parallel, we will also (for the first time) calculate the self-force by taking the $r \rightarrow r_p^+(t)$ limit in the mode-sum formula \eqref{eq:mode_sum_formula}, using the Gegenbauer procedure to compute the field modes in the external region. Having demonstrated this capability, our scalar-field code can be naturally modified to solve the Teukolsky equation instead, and thus compute the gravitational self-force using two-sided mode-sum regularization within the standard radiation gauge formalism \cite{Pound:2013faa}. Alternatively, given the solutions to the Teukolsky equation, we may also apply the recent Lorenz gauge metric reconstruction approach developed in Refs.~\cite{Dolan:2021ijg, Dolan:2023enf, Wardell:2024yoi}. In either case, an extension to a rotating Kerr background should be straightforward.

\begin{acknowledgments}
C.W. acknowledges support from the STFC via Grant No.~ST/Y00423X/1. L.B. acknowledges support from the STFC via Grant No.~ST/B001170/1. Computational resources from the University of Birmingham’s BlueBEAR High Performance Computing facility and the HPC system Viper at the Max Planck Computing and Data Facility were used in the preparation of this work.
\end{acknowledgments}

%%%%%%%%%%%%%%%%%%%%%%%%%%%%%%%%%%%%%%%%%%%
% Appendices
%%%%%%%%%%%%%%%%%%%%%%%%%%%%%%%%%%%%%%%%%%%%
\appendix
\section{Gegenbauer polynomials}\label{app:gegenbauer_polys}
The Gegenbauer polynomials $C_n^\lambda(s)$ are orthogonal with respect to the weighted inner product
\begin{align}
    \int_{-1}^{+1}(1-s^2)^{\lambda - 1/2}C_n^\lambda(s)C_m^\lambda(s) ds = h_n^\lambda \delta_{nm},
\end{align}
for $\lambda > -1/2$, where we adopt the normalization
\begin{align}
    h_n^\lambda = \pi^{1/2} C_n^\lambda(1)\frac{\Gamma\left(\lambda + \frac{1}{2}\right)}{(n+\lambda)\>\Gamma\left(\lambda\right)},\label{eq:GegC_normalization0}
\end{align}
with
\begin{align}
    C_n^\lambda(1) = \frac{\Gamma(n+2\lambda)}{n!\> \Gamma(2\lambda)}, \label{eq:GegC_normalization}
\end{align}
in line with Ref.~\cite{gottlieb1997gibbs}.  For $\lambda = 0$ the normalization is defined by the (well-defined) $\lambda \rightarrow 0$ limit of Eqs.~\eqref{eq:GegC_normalization0} and \eqref{eq:GegC_normalization}. 

Gegenbauer polynomials reduce to the standard Legendre polynomials when $\lambda = 1/2$: $C_n^{1/2}(s) = P_n(s)$. For $\lambda = 1$ they reduce to the Chebyshev polynomials of the second kind, $C_n^1(s) = U_n(s)$. For $\lambda = 0$, the normalization chosen in Eq.~\eqref{eq:GegC_normalization} causes $C_{n}^0(s) = 0$ for all $n > 0$, but alternative normalizations produce polynomials that are proportional to the Chebyshev polynomials of the first kind, $T_n(s)$.

\section{Numerical method}\label{app:numerical_methods}
In this appendix we summarize the numerical methods used for the scalar-field calculations of Sec.~\ref{sec:scalar_field}, describing the computation of the frequency-domain modes and our numerical implementation of the direct, EHS, and Gegenbauer reconstruction procedures. 

\subsection{Frequency-mode calculation}
\label{app:fd_calc_method}
The frequency-domain EHS modes $\tilde\psi^-_{\lmw}(r)$, defined in Eq.~\eqref{eq:int_ehs_def},
are calculated using the \texttt{C} language, EHS-based code developed in Ref.~\cite{Whittall:2023xjp}. To be precise, in the language of that paper we use the \texttt{IBP04corr6} method along with the \texttt{QAWO} adaptive quadrature routine from the GNU Scientific Library \cite{GSL_lib} to evaluate the normalization integrals $C_{\lmw}^-$ at frequencies $M|\omega| < 0.05$, and \texttt{IBP4corr6} at frequencies greater than this. The naming of these two methods indicates the different approaches they take to minimize the error committed by truncating Eq.~\eqref{eq:Cminus_def} at finite upper radius $\rmax$ (taken to be $2000M$ in this paper). See Sec. VB of Ref.~\cite{Whittall:2023xjp} for a description of these methods.

The inhomogeneous frequency-domain modes $\psi_{\lmw}(r)$ are calculated using a modified version of the EHS code. In particular, the second integral on the right-hand side of Eq.~\eqref{eq:VoPRadialEqnScattering},
\begin{align}
    	c_{\lmw}^-(r) := \displaystyle\int_{r}^{+\infty}\frac{\psi_{\lw}^+(r')S_{\lm\omega}(r')}{W_{\lw}}\frac{dr'}{f(r')}, 
\end{align}
differs from $C_{\lmw}^-$ only in the value of the lower limit, and so we evaluate these integrals using the same \texttt{IBP04corr6} algorithm. The first integral on the right-hand side of Eq.~\eqref{eq:VoPRadialEqnScattering},
\begin{align}
    	c_{\lmw}^+(r) := \displaystyle\int_{\rmin}^{r}\frac{\psi_{\lw}^-(r')S_{\lm\omega}(r')}{W_{\lw}}\frac{dr'}{f(r')}, 
\end{align}
differs from $C_{\lmw}^-$ both in the upper limit of integration and in the homogeneous solution that appears in the integrand [$\psi_{\lw}^-(r')$ rather than $\psi_{\lw}^+(r')$]. The $c_{\lmw}^+(r)$ integrals are evaluated using a modified version of the \texttt{IBP0corr0} routine from Ref.~\cite{Whittall:2023xjp} that incorporates the different limits and integrand. Note that the upper integration limit for $c_{\lmw}^+(r)$ is always finite, so we do not incur any error from artificial large-radius truncation in this case.

\begin{figure}[htb]
  \centering
  \includegraphics[width=\linewidth]{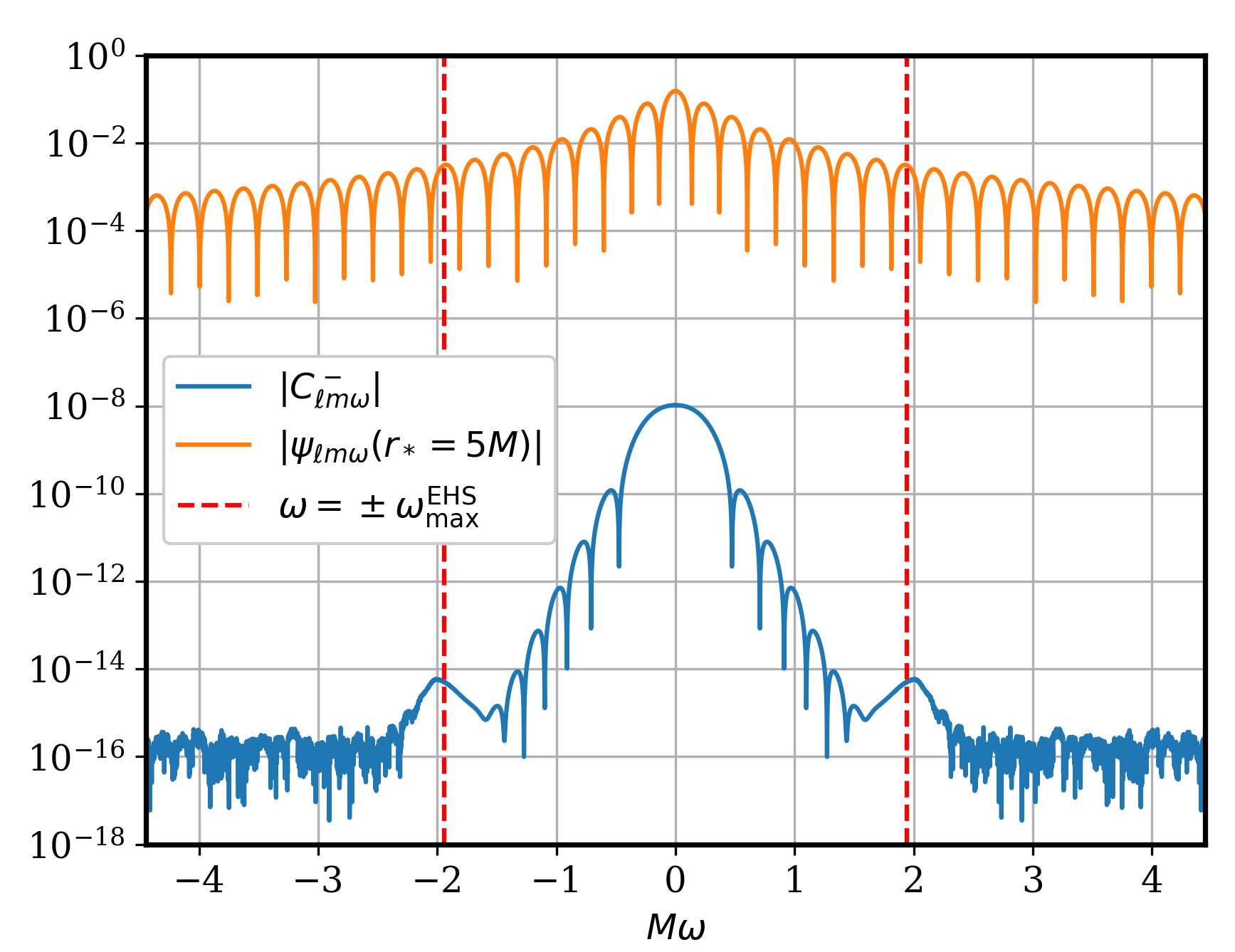}~\\~\\
  \caption{\label{fig:l10m0_FD_illustration}Illustration of the EHS normalization integral $C_{\lmw}^-$ and inhomogeneous field $\psi_{\lmw}(r_*=5M)$ as functions of $\omega$ for $(\ell, m) = (10, 0)$, obtained using the numerical method described in this appendix and derived from that of Ref.~\cite{Whittall:2023xjp}. Note the presence of noise in $C_{\lmw}^-$ at high frequencies---the numerical method automatically detects this noise, and truncates the EHS calculation accordingly, in this example at $|\omega| = \omega_{\rm max}^{\rm EHS} = 1.945/M$ (dashed red vertical lines). There is no such issue for the inhomogeneous field $\psi_{\lmw}$, which we may calculate to higher frequencies.}
\end{figure}

The values of the modes $\psi_{\lmw}(r)$ and $\tilde\psi^-_{\lmw}(r)$ are computed and stored at discrete frequencies (with spacing $M\Delta\omega = 1.25\times 10^{-3}$) for the different values of $r_*$ that we require. The frequencies for which we can calculate the EHS modes $\tilde\psi_{\lmw}^-$ are limited by the onset of noise in the calculated values of $C^-_{\lmw}$ at large values of $\omega$, as illustrated in Fig.~\ref{fig:l10m0_FD_illustration}. As described in Sec.~V D of Ref.~\cite{Whittall:2023xjp}, the code automatically detects the onset of noise and truncates the EHS calculation accordingly, returning the value of the truncation frequency, $\omega_{\rm max}^{\rm EHS}$. However, as can also be seen in Fig.~\ref{fig:l10m0_FD_illustration}, the calculation of the inhomogeneous modes $\psi_{\lmw}$ can be continued to larger frequencies without issue. Indeed, noise was not observed in the $\psi_{\lmw}$ spectra in any part of the numerical work conducted in this paper. Details of the frequency limits used for the inhomogeneous field in our numerical tests may be found in Sec.~\ref{sec:scalar_field}.

\subsection{Direct, EHS, and Gegenbauer reconstruction}

The direct, EHS, and Gegenbauer reconstruction approaches are implemented in the \texttt{Python} programming language. Given pregenerated data for the frequency-domain quantities $\psi_{\lmw}(r)$, $C_{\lmw}^-$, and $\psi_{\lw}^-(r)$ on their discrete frequency grids, cubic spline interpolants over frequency are constructed using the \texttt{InterpolatedUnivariateSpline} routine from the \texttt{scipy} library \cite{2020SciPy-NMeth}. The direct reconstruction of the time-domain spherical harmonic mode $\psi_{\lm}(t,r)$ and its time derivative $\partial_t\psi_{\lm}(t,r)$ is achieved by numerically evaluating the partial Fourier integrals
\begin{align}
    \psi_{\lm}(t,r) &\approx \displaystyle\int_{-\wmax^{\rm VoP}}^{+\wmax^{\rm VoP}} \psi_{\lmw}(r) e^{-i\omega t} d\omega,\label{eq:field_partial}\\
    \partial_t\psi_{\lm}(t,r) &\approx -\displaystyle\int_{-\wmax^{\rm VoP}}^{+\wmax^{\rm VoP}} i\omega \>\psi_{\lmw}(r) e^{-i\omega t} d\omega,\label{eq:deriv_partial}
\end{align}
where $\wmax^{\rm VoP}$ is the chosen truncation frequency. Likewise, for the EHS reconstruction we evaluate
\begin{align}
    \tilde\psi_{\lm}^-(t,r) &\approx \displaystyle\int_{-\wmax^{\rm EHS}}^{+\wmax^{\rm EHS}} C_{\lmw}^-\psi_{\lw}^-(r) e^{-i\omega t} d\omega,\\
    \partial_t\tilde\psi_{\lm}^-(t,r) &\approx -\displaystyle\int_{-\wmax^{\rm EHS}}^{+\wmax^{\rm EHS}} i\omega \>C_{\lmw}^-\psi_{\lw}^-(r) e^{-i\omega t} d\omega. \label{eq:ehs_deriv_partial}
\end{align}
Integrals \eqref{eq:field_partial} -- \eqref{eq:ehs_deriv_partial} are evaluated using the adaptive Gauss-Kronrod routine included in the \texttt{quad} integrator from \texttt{scipy}. The Gegenbauer reconstruction is achieved by first performing the direct reconstruction procedure, evaluating integrals \eqref{eq:field_partial} and \eqref{eq:deriv_partial} as above (storing these fields on time grids of spacing $\Delta t = 0.04M$ and constructing cubic spline interpolants), and then using \texttt{quad} to numerically evaluate the integral \eqref{eq:g_k_def} for the Gegenbauer expansion coefficients, $g_k^\lambda$. The Gegenbauer approximant \eqref{eq:geg_approximant_example} is computed using the \texttt{scipy} routine \texttt{eval\_gegenbauer} to evaluate the polynomials $C_k^{\lambda}$.

%%%%%%%%%%%%%%%%%%%%%%%%%%%%%%%%%%%%%%%
\bibliographystyle{unsrt}
\bibliography{references}
%%%%%%%%%%%%%%%%%%%%%%%%%%%%%%%%%%%%%%%

\end{document}